%% file: main.tex
\documentclass[aps,prr,twocolumn,groupedaddress]{revtex4-2}

\usepackage{amsmath}
\usepackage{amssymb}
\usepackage{accents}
\usepackage{contour}
\usepackage{relsize}
\usepackage{mathtools}
\usepackage{dsfont}
\usepackage{natbib}
\usepackage[hidelinks]{hyperref}
\usepackage{tikz}
\usepackage[bb=stix,bfbb]{mathalpha}
\usepackage{bm}

\usepackage{graphicx}
\usepackage{placeins}
\usepackage{subcaption}

\newcommand*{\dt}[1]{%
	\accentset{\mbox{\large\bfseries .}}{#1}}

\newcommand{\obliquesymbol}[1]{%
	\begin{tikzpicture}[#1]%
		\draw (0,0) -- (1.0ex,0);%
		\draw (0.5ex,0) -- (1.0ex,0.8ex);%
	\end{tikzpicture}%
}

\graphicspath{{imgs/}{imgs2/}}

\newcommand{\leqnos}{\tagsleft@true\let\veqno\@@leqno}
\newcommand{\reqnos}{\tagsleft@false\let\veqno\@@eqno}

\newcommand{\vfield}[1]{
	\boldsymbol{#1}
}
\newcommand{\pfield}[1]{
	\mathrm{#1}
}
\newcommand{\vpfield}[1]{
	\mathbb{#1}
}
\newcommand{\efield}[1]{
	\mathbb{\widehat{#1}}
}
\newcommand{\inv}[1]{
	\mathrm{#1^{-1}}
}
\newcommand{\mathsfbi}[1]{
	\bm{\mathsf{#1}}
}

\def\vecU{\boldsymbol{U}}
\def\vecUb{\vfield{U^{0}}}

\def\vecu{\vfield{u}}
\def\vecv{\vfield{v}}

\def\vecyp{\mathbb{y}}
\def\bcdot{\boldsymbol{\cdot}}
\def\ie{\textit{i.e.} }
\def\blbracket{\boldsymbol{\lceil}}
\def\brbracket{\boldsymbol{\rceil}}

\def\invRec{\inv{Re_{c}}}
\def\Rey{\mathrm{Re}}

\DeclareBoldMathCommand\bigblbracket{\left\lceil}
\DeclareBoldMathCommand\bigbrbracket{\right\rceil}

\begin{document}

\title{Asymptotic Center--Manifold for the Navier--Stokes}
\author{Prabal S. Negi}
\email[]{prabal.negi@oist.jp}
\affiliation{Okinawa Institute of Science and Technology Graduate University, Onna, Okinawa 904-0495, Japan}

\date{\today}

\begin{abstract}
Center-manifold approximations for infinite-dimensional systems are treated in the context of the Navier--Stokes equations extended to include an equation for the parameter evolution. The consequences of system extension are non-trivial and are examined in detail. The extended system is reformulated via an isomorphic transformation, and the application of the center-manifold theorem to the reformulated system results in a finite set of center-manifold amplitude equations coupled with an infinite-dimensional graph equation for the stable subspace solution. General expressions for the asymptotic solution of the graph equation are then derived. The main benefit of such an approach is that the graph equation, and the subsequent asymptotic expressions are formally valid even when the system is perturbed slightly away from the bifurcation point. The derivation is then applied to two cases - the classic case of the Hopf bifurcation of the cylinder wake, and a case of flow in an open cavity which has interesting dynamical properties after bifurcation. Predictions of the angular frequencies of the reduced systems are in good agreement with those obtained for the full systems close to the bifurcation point. The Stuart-Landau equations for the two cases are also obtained. The presented methodology may easily be applied to other infinite-dimensional systems. 
\end{abstract}

\maketitle

\input{intro}

\input{derivation_p1}

\input{cyl}

\input{cavity}

\input{conclusion}
\input{appendix}

\FloatBarrier

\bibliographystyle{apalike}
\bibliography{references}

\end{document}

%% file: intro.tex
\section{Introduction}

Center-manifold theory provides an elegant and powerful tool for the analysis of dynamics of systems in the vicinity of the bifurcation point \citep{carr82,carr83b,sijbrand85,guckenheimer83,wiggins03}. In such systems the essential dynamics are largely driven by the dynamics in the center-manifold, which is a continuation of the center space of the linearized system into the non-linear regime. Being a continuation of the center space, the center manifold has the same dimension as the center space and leads to a low-dimensional representation of what might originally be a very high (or infinite) dimensional system. The ideas underlying the theory are powerful and extend beyond center manifolds, having found applicability to general invariant manifolds \citep{roberts89}. In the context of hydrodynamics, the investigations of such systems is often done by invoking the idea of separation of rapid oscillations from slowly modulating amplitudes \citep{stuart60,watson60,newell69}. In the more recent literature it is often referred to as weakly non-linear analysis and follows the formal approach of multiple-scale analysis \citep{sipp07,meliga11,meliga12}. The procedure results in a set of equations for the slowly varying part of the amplitudes of the unstable eigenmode(s), referred to as the amplitude equations \citep{cross09} or sometimes the Stuart-Landau equations. While it has been claimed that the two methods are effectively equivalent \citep{fujimara91}, the underlying ideas of the two methods are somewhat distinct, leading to contrasting procedures for the formal evaluation of the approximations. As opposed to the method of multiple-scales, the center manifold theory assumes no separation of time scales and divides the solution space into a ``dynamic'' component corresponding to center space of the linearized system at bifurcation and a ``driven'' component corresponding to the stable subspace. The center-manifold theorem leads to a graph equation for the driven component of the solution. While the application of center-manifold theory to ordinary differential equation type problems is fairly straightforward the application to partial differential equations, which are infinite dimensional, poses a harder challenge since the graph equation is a highly non-linear, infinite-dimensional problem. 

In the literature the approach to the problem appears to converge to computing the normal form of the reduced center-manifold system \citep{knobloch83,guckenheimer83b,coullet83,haragus11}. Differences arise in the flavor of the approach in the various studies. For example simple self-adjoint problems were investigated in \cite{carr83} where the solution to the graph equation was shown to vanish at second order, resulting in simplified evaluation of the center manifold. Sometimes reasonable justification may be provided for truncating the stable subspace to finite dimension \citep{knobloch83,guckenheimer83b,fujimura97}, in which case the center-manifold approximation can effectively be treated in manner similar to ordinary differential equations. An iterative numerical approach has been pursued in \cite{roberts97} which sought to minimize the residual of the amplitude equation for the critical modes to determine the total solution. A similar algorithmic approach was undertaken in \cite{carini15} for the approximation of normal forms for the center manifolds. The variations not withstanding, the main idea commonly utilized is to approximate the non-linear transforms required for achieving the normal form for the low-dimensional (center-manifold) system, with Taylor expansions used to treat variation of the system parameters \citep{haragus11}. 

The current work takes a different approach for the evaluation of the center-manifold, falling back to the spirit of its computation for ordinary differential equations \citep{wiggins03}, while also accounting for the infinite-dimensional nature of the problem. This has been done in the context of the Navier--Stokes however, the underlying idea may easily be adapted to other systems. Taking a step back, the Navier--Stokes system is first extended to include the trivial equation for the parameter evolution, with the parameter now treated as a dynamic variable. This increases the dimension of the center-manifold which has non-trivial consequences as shall be seen. This requires understanding how the spectral properties of the extended system relate to those of the original problem, which is done in a general setting. The system is then transformed into an appropriate form and the center-manifold theorem is then applied to this modifed problem leading to an infinite-dimensional graph equation, which is solved asymptotically. The graph equation and the subsequent asymptotic solutions are formally valid for systems that have been perturbed away from the bifurcation point, as opposed to the case of the center-manifold reduction of the standard problem where the graph equation is valid only at the bifurcation point. A major consequence of this approach is that when the system is perturbed away from the bifurcation point one does not need to consider the Taylor expansions with respect to the system parameter(s). The solution is essentially baked in to the asymptotic expressions that are derived. The primary focus of the work then is not so much on the derivation of the normal forms but rather in the evaluation of the graph equations. Once the solutions to the graph equations are known, the subsequent step of obtaining the ``amplitude equations'' is obtained almost as a byproduct. In this sense, the current work may also be viewed as an alternate route for a rigorous derivation of the Landau model \citep{landau52} in infinite-dimensional systems, which is often used to understand the dynamics of weakly non-linear systems. 

The remainder of the manuscript is laid out as follows. Section~\ref{problem_setup} first describes the notation used in the current manuscript, then presents the problem in a general setting of the Navier--Stokes equation and subsequently explores the consequences of extending the system with the trivial parameter evolution equation. Section~\ref{center_manifold_derivation} then reformulates the problem in an appropriate form for the application of center-manifold theory. An asymptotic approximation is then carried out and the resulting expressions for the power series are obtained. The presented theory is then applied to the case of a Hopf bifurcation in a cylinder wake in section~\ref{application_cylinder} and to the case of the flow in an open cavity in section~\ref{application_cavity}. Concluding remarks are made in section~\ref{conclusion}.

%% file: derivation_p1.tex
\section{Problem Setup}
\label{problem_setup}

\subsection{Notation}
In the subsequent sections a number of mathematical objects, with varying definitions of the space to which they belong (or act on), are encountered. An effort has been made to make the notation compact where possible so that the essential meaning of an expression or operation is clear without the need to write out all individual terms repeatedly. Therefore in this subsection some compact notations and the general rule that has been used for the definition of new objects is described.

Throughout the text, bold symbols ($\vecUb,\vecu,\vecu_{s},\vecv,$ \textit{etc.}) will refer to the velocity fields for the case in consideration. The velocity fields will have an an associated pressure field which ensures the divergence-free evolution of the velocities, which is denoted in Roman font ($\pfield{P^{0}}, \pfield{p},$ \textit{etc.}). The two fields will often appear together in equations, so the pair is represented compactly using the Blackboard bold font as $\mathbb{u} = (\vfield{u},\pfield{p})$, retaining the same alphabet as the velocity field. Later the original system is extended by treating the parameter as a dynamic variable. The extended state variable is then represented by including a hat on the compact notation as $\widehat{\mathbb{u}} = (\mathbb{u},\eta) = (\vfield{u},\pfield{p},\eta)$. Operators are represented using caligraphic font ($\mathcal{L}$) and adjoint operators and variables are indicated by a $^{\dagger}$ symbol. The ``hat'' notation is used for other mathematical objects which are associated with the extended space, so that a linear operator acting on the extended vector space is written as $\mathcal{\widehat{L}}$. Inner products are denoted by $\langle\cdot,\cdot\rangle$ and the vector space underlying the inner-product is inferred from the specific pair of vectors that the inner-product is acting on. Matrices are defined using bold sans serif font ($\mathsfbi{V},\mathsfbi{\widehat{W}},$ \textit{etc}). 
Since the momentum and divergence equations are written compactly as one using the Blackboard bold notation, a situation will often be encountered where certain terms appear in the momentum equations but have no counter-part in the equation for the divergence-free constraint. Ceiling brackets, $\blbracket \bcdot \brbracket$, are used for such terms. So the non-linear term of the Navier--Stokes will be denoted as
\begin{equation}
	\blbracket \vfield{u}\bcdot\nabla\vfield{u} \brbracket =
	\begin{Bmatrix}
		\vfield{u}\bcdot\nabla\vfield{u} \\
		\pfield{0}
	\end{Bmatrix}, \nonumber
\end{equation}
signifying the contribution of the term to the momentum equations along with a null contribution to the continuity equation. Subsequently the extended dynamical system will include the parameter as a dynamic variable. One again encounters scenarios where certain terms have no counter part for this extended equation. The ceiling bracket notation will be used recursively so that quantities of the type $\blbracket \blbracket \vfield{u} \brbracket\brbracket$ and $\blbracket\vpfield{u}\brbracket$ have the following meaning,
\begin{equation}
	\begin{aligned}
		\blbracket \blbracket \vfield{u} \brbracket\brbracket =
		\blbracket (\vfield{u},\pfield{0}) \brbracket =
		\begin{Bmatrix}
			\vfield{u} \\
			\pfield{0} \\
			0
		\end{Bmatrix}; 
		&	&
		\blbracket \vpfield{u} \brbracket =
		\begin{Bmatrix}
			\vfield{u} \\
			\pfield{p} \\
			0
		\end{Bmatrix}.
	\end{aligned}
	\nonumber
\end{equation}
Additionally, two matrices $\mathsfbi{B}$ and $\mathsfbi{\widehat{B}}$ are defined as 
\begin{equation}
	\begin{aligned}
		\mathsfbi{B} =
		\begin{bmatrix}
			1 & \pfield{0} \\ 
			0 & \pfield{0}
		\end{bmatrix}; 
		&	&
		\mathsfbi{\widehat{B}} =
		\begin{bmatrix}
			1 & 0 & 0 \\ 
			0 & 0 & 0 \\
			0 & 0 & 1
		\end{bmatrix}
	\end{aligned}
	\nonumber,
\end{equation}
so that the time derivatives are written compactly as 
\begin{equation}
	\begin{aligned}
		\dfrac{\partial (\mathsfbi{B}\vpfield{u})}{\partial t} =
		\begin{Bmatrix}
			\dfrac{\partial \vfield{u}}{\partial t} \vspace{1mm}\\
			\pfield{0}
		\end{Bmatrix}; 
		&	&
		\dfrac{\partial (\mathsfbi{\widehat{B}}\efield{u})}{\partial t} =
		\begin{Bmatrix}
			\dfrac{\partial \vfield{u}}{\partial t} \vspace{1mm}\\
			\pfield{0} \vspace{1mm}\\
			\dfrac{\partial \eta}{\partial t}
		\end{Bmatrix}
	\end{aligned},
	\nonumber
\end{equation}
resulting from the special nature of the incompressibility equation.

\subsection{The Standard Problem}
Consider the case of of a flow $\vpfield{U} = (\vfield{U},\pfield{P})$ governed by the Navier--Stokes equation, with an appropriate set of boundary conditions, at a specified Reynolds number $\Rey$, given by 
\begin{equation}
	\label{NavierStokes}
	\begin{aligned}
		\frac{\partial \vfield{U}}{\partial t} =& - \vecU\bcdot\nabla \vecU - \nabla \pfield{P} + \inv{Re}\nabla^{2} \vecU, \\
		0 =& \nabla\bcdot \vecU.
	\end{aligned}
\end{equation}
It is presumed that the system has a fixed point $\vpfield{U^{0}} = (\vecUb,\pfield{P^{0}})$ for a critical parameter value of $Re_{c}$, at which the system undergoes a bifurcation. The state $\vpfield{U^{0}}$ satisfies the stationary Navier-Stokes equation,
\begin{equation}
	\label{NS_stationary}
	\begin{aligned}
		-\vecUb\bcdot\nabla \vecUb - \nabla \pfield{P^{0}} + \inv{Re_{c}}\nabla^{2} \vecUb = 0, \\
		\nabla\bcdot \vecUb	= 0.
	\end{aligned}
\end{equation}
Decomposing an instantaneous flow field as a deviation from the stationary state, $\vpfield{U} = \vpfield{U}^{0} + \vpfield{u}$, given by
\begin{equation}
\begin{aligned}
	\label{flow_field_decomposition}
	\vecU 			   = \vecUb + \vecu, && 
	\pfield{P} 		= \pfield{P^{0}} + \pfield{p},
\end{aligned} \nonumber
\end{equation}
one obtains the evolution equations for deviations $\mathbb{u} = (\vecu,\pfield{p})$ as 
\begin{eqnarray}
	\label{NS_deviation}
	\begin{aligned}
		\frac{\partial \vecu}{\partial t} =& -\vecUb\bcdot\nabla \vecu - \vecu\bcdot\nabla\vecUb - \vecu\bcdot\nabla\vecu - \nabla \pfield{p} + \invRec\nabla^{2}\vecu,&& \\
		0 =& \nabla\bcdot \vecu. &&
	\end{aligned}
\end{eqnarray}
Linearizing the flow near the stationary state $\mathbb{U^{0}}$ one obtains the first order equations for the evolution of the perturbations as
\begin{equation}
	\label{NS_perturbation}
	\begin{aligned}
		\frac{\partial \vecu}{\partial t} =& -\vecUb\bcdot\nabla \vecu - \vecu\bcdot\nabla\vecUb - \nabla \pfield{p} + \inv{Re_{c}}\nabla^{2} \vecu, \\
		0 =& \nabla\bcdot \vecu .
	\end{aligned}
\end{equation}
The standard linearized operator at the bifurcation point acting on $\mathbb{u}$ is denoted as
\begin{eqnarray}
	\mathcal{L}(\vpfield{u}) =
	\begin{bmatrix}
			-\vecUb\bcdot\nabla  - (\nabla\vecUb)\bcdot + \inv{Re_{c}}\nabla^{2} & -\nabla \\
			\nabla \bcdot  													     										  & 0
	\end{bmatrix}
	\begin{Bmatrix}
		\vecu \\
		\pfield{p}
	\end{Bmatrix},
	\label{Lop}
\end{eqnarray}
with the full non-linear problem written compactly as
\begin{eqnarray}
	\label{NS_deviation_compact}
	\frac{\partial (\mathsfbi{B}\vpfield{u}) }{\partial t} =  \mathcal{L}(\mathbb{u}) - \blbracket \vecu\bcdot\nabla \vecu \brbracket \label{nonlinear_evolution}.
\end{eqnarray}
In addition, the adjoint operator to $\mathcal{L}$ is denoted as $\mathcal{L}^{\dagger}$. It is assumed that the matrix pair $(\mathcal{L}, \mathsfbi{B})$ has eigenpairs $\{\lambda_{i},\mathbb{v}_{i}\}$, where, $\lambda_{i}$ is the eigenvalue and $\mathbb{v}_{i}$ is the corresponding right eigenvector. Similarly, $(\mathcal{L},\mathsfbi{B})^{\dagger}$ has eigenpairs $\{\lambda^{*}_{i},\mathbb{v}^{\dagger}_{i}\}$, where $^{*}$ denotes the complex-conjugtation. The scaling of eigenvectors is chosen to satisfy the biorthogonality condition
\begin{eqnarray}
	\langle\mathbb{v}^{\dagger}_{i},\mathsfbi{B}\mathbb{v}_{j}\rangle = \delta_{i,j} \label{std_biorthogonality}.
\end{eqnarray}
The solutions to equation~\eqref{NS_deviation_compact} are assumed to lie in a Hilbert space $\mathds{H}$. The bifurcation of the system at the critical Reynolds number $Re_{c}$ implies that one or more of the eigenmodes of the linearized operator represented by the matrix pair $(\mathcal{L},\mathsfbi{B})$ has eigenvalues $\lambda$ with $\mathfrak{Re}(\lambda) = 0$, herein referred to as the critical eigenvalues. The space spanned by the eigenvectors of the critical eigenvalues is referred to as the center space of $(\mathcal{L},\mathsfbi{B})$. In the context of center-manifold approximation it is sometimes also referred to as the critical subspace. Accordingly, the Hilbert space $\mathds{H}$ may be decomposed as the direct sum of two invariant subspaces such that
\begin{equation}
	\label{Hilbert_space_division}
	\begin{aligned}
		\mathds{H} =& \mathds{T}_{c}\bigoplus \mathds{T}_{s},& \\
		\forall \ \mathbb{u} \in& \mathds{T}_{c},& \mathcal{L}(\mathbb{u}) \in \mathds{T}_{c},  \\
		\forall \ \mathbb{u} \in& \mathds{T}_{s},& \mathcal{L}(\mathbb{u}) \in \mathds{T}_{s}.  
	\end{aligned} 
\end{equation}
Here, it has been implicitly assumed that $(\mathcal{L},\mathsfbi{B})$ has no unstable eigenvalues. $\mathds{T}_{c}$ denotes the center subspace of $(\mathcal{L},\mathsfbi{B})$ and $\mathds{T}_{s}$ will be referred to as the stable subspace. A matrix $\mathsfbi{V}$ is defined, the columns of which comprise of the eigenvectors corresponding to the critical eigenvalues of $(\mathcal{L},\mathsfbi{B})$, and $\mathsfbi{W}$ is defined as the corresponding matrix for the adjoint operator $(\mathcal{L},\mathsfbi{B})^{\dagger}$. Due to the chosen scaling one obtains the relation $\mathsfbi{W}^{H}\mathsfbi{B}\mathsfbi{V} = \mathsfbi{I}_{n}$, where, $n$ is the dimension of the center subspace and $\mathsfbi{I}_{n}$ is an identity operator of order $n$. 

\subsection{The Extended Problem}
Now consider the full nonlinear evolution equation of the flow field at a parameter value $\inv{Re} = \invRec(1  - \eta)$ where, the deviation of the inverse Reynolds number from its critical value is given by $-\invRec\eta$. The evolution equations for the (deviation) velocity and pressure fields are given by
\begin{subequations}
	\label{NS_CM_0}
	\begin{eqnarray}
		\label{NS_2}
			\frac{\partial \vfield{u} }{\partial t} =& \left\lbrace
			\begin{split}
				\invRec \nabla^{2}\vecu - \vecUb\bcdot\nabla \vecu - \vecu\bcdot\nabla\vecUb \\
				- \nabla \pfield{p}
				- \invRec\eta \nabla^{2}\vecUb \\
				- \vecu\bcdot\nabla \vecu 
				-\invRec\eta\nabla^{2}\vecu
			\end{split}\right. \\
			0 =& \nabla\bcdot \vfield{u}, \\
		\label{parameter_evolution}
			\frac{\partial \eta}{\partial t} =& 0,
		\end{eqnarray}
\end{subequations}
where the parameter perturbation is included as an additional dynamical equation, in the spirit of its treatment for ordinary differential equations \citep{wiggins03,guckenheimer83} for center-manifold approximations. This system with an additional dynamical equation for the parameter will be referred to as the \emph{extended system}. The new linearized operator $\mathcal{\widehat{L}}$ acting on the extended state $\mathbb{\widehat{u}} = (\mathbb{u},\eta)$, is represented by
\begin{equation}
	\mathcal{\widehat{L}}(\mathbb{\widehat{u}}) =
	\begin{bmatrix}
		\mathcal{L}& -\blbracket\inv{Re_{c}}\nabla^{2}\vecUb \brbracket \\
		0 	& 0
	\end{bmatrix}
	\begin{Bmatrix}
		\mathbb{u} \\
		\eta
	\end{Bmatrix}
	\label{Lop_CM},
\end{equation}
which is referred to as the \emph{linearized extended system}. One may write the full nonlinear system  represented by equation~\eqref{NS_CM_0} in a more compact form as
\begin{equation}
	\label{NS_CM_0_compact}
	\dfrac{\partial (\mathsfbi{\widehat{B}}\efield{u}) }{\partial t} =
	\mathcal{\widehat{L}}(\efield{u})  -  \bigblbracket \bigblbracket \invRec\eta\nabla^{2}\vecu \bigbrbracket\bigbrbracket 
	- \blbracket\blbracket \vecu\bcdot\nabla \vecu \brbracket\brbracket , 
	%
\end{equation}
with the terms in the ceiling brackets denoting the non-linear terms. 

A few observations are made regarding the spectral properties of $(\mathcal{\widehat{L}},\mathsfbi{\widehat{B}})$. If $\{\lambda,\mathbb{v}\}$ is an eigenpair of $(\mathcal{L},\mathsfbi{B})$ then, clearly $\{\lambda,(\mathbb{v},0)\}$ is an eigenpair of the linearized extended system $(\mathcal{\widehat{L}},\mathsfbi{\widehat{B}})$ (which may be verified via substitution). Thus all eigenpairs of the standard linear problem can be trivially extend to the linearized extended problem where the additional dynamic variable $\eta$ vanishes.

The linearized extended system contains another eigenpair which is not present in the original system, $\{\lambda_{0},(\mathbb{v}_{0},\zeta_{0})\}$, where $\lambda_{0} = 0$ and the eigenvector components are not as trivial. After substitution into the eigenvalue equation for $\lambda = 0$ one obtains
\begin{eqnarray}
	\mathcal{L}(\vpfield{v}_{0})  -
	\zeta_{0}\blbracket \invRec\nabla^{2}\vecUb \brbracket = 0, 
	\label{extended_eigenvector}
\end{eqnarray}
where $\zeta_{0}$ is used to scale the eigenvector such that it satifies the appropriate normalization condition. Thus one obtains a new eigenpair for the extended system with a trivial eigenvalue and an eigenvector, $\mathbb{\widehat{v}}_{0} = (\mathbb{v}_{0},\zeta_{0})$, which satisfies equation~\eqref{extended_eigenvector}. This new mode will be referred to as the ``parameter mode''.

 The adjoint operator corresponding to $\mathcal{\widehat{L}}$ is obtained as 
\begin{equation}
	\mathcal{\widehat{L}^{\dagger}}(\mathbb{\widehat{u}}^{\dagger}) = 
	\begin{bmatrix}
		\mathcal{L}^{\dagger}& 0 \\
		-\blbracket \invRec\nabla^{2}\vecUb) \brbracket^{H} 	& 0
	\end{bmatrix}
	\begin{Bmatrix}
		\mathbb{u}^{\dagger} \\
		\eta^{\dagger}
	\end{Bmatrix}
	\label{Adjoint_Lop_CM},
\end{equation} 
where, the superscript $^{H}$ denotes the complex-conjugated transpose and
\begin{eqnarray}
	\blbracket \invRec\nabla^{2}\vecUb \brbracket^{H}(\mathbb{u}^{\dagger}) = \langle\invRec\nabla^{2}\vecUb,\vecu^{\dagger}\rangle. \nonumber
\end{eqnarray}

For a properly defined adjoint, the spectra of the direct and adjoint problems must be consistent. The new ``adjoint parameter mode'' corresponding to $\lambda^{*}_{0} = 0$ is easily seen to be $\{\lambda^{*}_{0},(\mathbb{v}^{\dagger},\zeta^{\dagger})_{0}\} = \{0,(\mathbb{0},\zeta^{\dagger}_{0})\}$, where the value of $\zeta^{\dagger}_{0}$ depends on the chosen norm. For the adjoint modes that are extensions of the original linear problem one finds that if $\{\lambda^{*}_{i},\mathbb{v}^{\dagger}_{i}\}$ is an eigenpair for $(\mathcal{L},\mathsfbi{B})^{\dagger}$, the corresponding eigenpair for $(\mathcal{\widehat{L}},\mathsfbi{\widehat{B}})^{\dagger}$ is $\{\lambda^{*}_{i},(\mathbb{v}^{\dagger},\zeta^{\dagger})_{i}\}$, with $\zeta^{\dagger}_{i} = -\langle\invRec\nabla^{2}\vecUb,\vfield{v}^{\dagger}_{i}\rangle/\lambda^{*}_{i}$. Note that 
as $\lambda^{*}_{i}$ approaches zero the eigenvector of this mode approaches the adjoint parameter mode. One must then consider the Jordan or Weierstrass canonical form for the generalized eigenvectors. This particular degenerate case is not addressed in the current manuscript. The biorthogonality condition for the extended system results in
\begin{eqnarray}
	\langle\mathbb{\widehat{v}}^{\dagger}_{i},\mathsfbi{\widehat{B}}\mathbb{\widehat{v}}_{j}\rangle = \langle\vpfield{v}^{\dagger}_{i},\mathsfbi{B}\vpfield{v}_{j}\rangle + (\zeta^{\dagger}_{i})^{*}\zeta_{j} = \delta_{i,j}. \label{extended_biorthogonality}
\end{eqnarray}
For all the extended eigenmodes this reduces to equation~\eqref{std_biorthogonality} since $\zeta_{j} = 0$ for all $j \ge 1$. The new adjoint parameter mode has $\mathbb{v}^{\dagger}_{0} = 0$ which reduces the biorthogonality relation for the parameter mode to $(\zeta^{\dagger}_{0})^{*}\zeta_{0} = 1$. For the rest of the manuscript it is assumed that $\zeta^{\dagger}_{0} = \zeta_{0} = 1$, and $(\boldsymbol{v}_{0},p_{0})$ are given by equation~\eqref{extended_eigenvector} with $\zeta_{0} = 1$.

Hence, if one has calculated the eigendecomposition (partial or complete) of the original linear problem then one may easily obain the corresponding eigen-decomposition for the extended linear problem with an additional parameter mode that has $\lambda = 0$ as its eigenvalue.

As in the standard problem, one may assume the solutions $\efield{u}$ to lie in an extended Hilbert space $\mathds{\widehat{H}}$, which may be expressed as a direct sum of the two invariant center and stable subspaces of $(\mathcal{\widehat{L}},\mathsfbi{\widehat{B}})$ so that
\begin{equation}
	\begin{aligned}
		\mathds{\widehat{H}} =& \mathds{\widehat{T}}_{c}\bigoplus \mathds{\widehat{T}}_{s},& \\
		\forall \ \mathbb{\widehat{u}} \in& \mathds{\widehat{T}}_{c}, & \mathcal{\widehat{L}}(\mathbb{\widehat{u}}) \in \mathds{\widehat{T}}_{c}, \\
		\forall \ \mathbb{\widehat{u}} \in& \mathds{\widehat{T}}_{s}, & \mathcal{\widehat{L}}(\mathbb{\widehat{u}}) \in \mathds{\widehat{T}}_{s}. 
	\end{aligned} 
\end{equation}
One may build matrices $\mathsfbi{\widehat{V}}$ and $\mathsfbi{\widehat{W}}$ whose columns comprise of the critical eigenvectors of the direct and adjoint operators respectively, such that $\mathsfbi{\widehat{W}}^{H}\mathsfbi{\widehat{B}}\mathsfbi{\widehat{V}} = \mathsfbi{I}_{m}$. If $n$ is the dimension of the center subspace $\mathds{T}_{c}$, then clearly the dimension of $\mathds{\widehat{T}}_{c}$ is $m = n + 1$.

It is important to point out at this stage that in the linearized extended system, the only eigenvector that has a non-zero component in the dynamic parameter variable $\eta$ is the parameter mode, which belongs to the critical subspace $\mathds{\widehat{T}}_{c}$. Therefore in the stable subspace $\mathds{\widehat{T}}_{s}$ the extended variable $\eta$ vanishes identically. As an additional note, in the current work the only free parameter is the Reynolds number so the critical subspace dimension is extended only by one. However, one may apply the same procedure for multiple parameters resulting in several ``parameter modes'' for the extension of the critical subspace. 

\section{Center Manifold Approximation}
\label{center_manifold_derivation}

\subsection{Problem Reformulation}
The starting point of the center-manifold reduction is taken to be equation~\eqref{NS_CM_0} or its compact form \eqref{NS_CM_0_compact}, which is reformulated in this section so as to express it in a form appropriate for the application of the center-manifold theory. The system is assumed to have a critical subspace $\mathds{\widehat{T}}_{c}$ of dimension $m$ and that the rank-$m$ matrices $\mathsfbi{\widehat{V}}$ and $\mathsfbi{\widehat{W}}$ are defined. The critical eigenvectors of $\mathcal{\widehat{L}}$ which define the columns of $\mathsfbi{\widehat{V}}$ are numbered using a subscript as $\mathsfbi{\widehat{V}} = [\mathbb{\widehat{v}}_{0}, \mathbb{\widehat{v}}_{1}, \ldots, \mathbb{\widehat{v}}_{m-1}]$, with $\efield{v}_{0}$ being the parameter mode. Similarly $\mathsfbi{\widehat{W}} = [\mathbb{\widehat{v}}^{\dagger}_{0}, \mathbb{\widehat{v}}^{\dagger}_{1}, \ldots, \mathbb{\widehat{v}}^{\dagger}_{m-1}] $ represent the corresponding adjoint eigenvectors and, $\mathsfbi{\widehat{\Lambda}} = diag(\lambda_{0},\lambda_{1},\ldots,\lambda_{m-1})$ is the diagonal matrix of the critical eigenvalues of the extended system. Additionally, the matrices $\mathsfbi{V} = [\mathbb{v}_{1}, \ldots, \mathbb{v}_{m-1}]$ and $\mathsfbi{W} = [\mathbb{v}^{\dagger}_{1}, \ldots, \mathbb{v}^{\dagger}_{m-1}]$ are the corresponding matrices of the original linear system. This implicitly assumes that the linear operator is diagonalizable. In the case of defective matrices one would need to replace $\mathsfbi{\widehat{\Lambda}}$ with the equivalent Jordan form. The direct and adjoint eigenmodes would be replaced by the respective generalized eigenmodes and much of the theory would extend in a straightforward manner. 

The solution $\mathbb{\widehat{u}}$ of equation~\eqref{NS_CM_0_compact} is decomposed as the sum of its components in $\mathds{\widehat{T}}_{c}$ and $\mathds{\widehat{T}}_{s}$, \ie $\mathbb{\widehat{u}} = \mathbb{\widehat{u}}_{c} + \mathbb{\widehat{u}}_{s}$.  Since $\mathds{\widehat{T}}_{c}$ is a finite dimensional subspace, we may decompose $\mathbb{\widehat{u}}_{c}$ further into the individual components of each eigenvector so that 
\begin{eqnarray}
	\mathbb{\widehat{u}}_{c} = \sum_{i=0}^{m-1}x_{i}\mathbb{\widehat{v}}_{i} = \mathsfbi{\widehat{V}}\mathsfbi{x},
	& \hspace{5mm}
	\mathbb{\widehat{u}} = \mathsfbi{\widehat{V}}\mathsfbi{x} + \mathbb{\widehat{u}}_{s} \label{solution_decomposition},
\end{eqnarray}
where the $x_{i}$'s are the scalar time-dependent (complex) amplitudes of the critical eigenvectors. The ordered collection of the scalars is represented as a vector $\mathsfbi{x} = [x_{0}, x_{1},\ldots x_{m-1}]^{T}$. Due to the chosen scaling of the eigenvectors, the parameter value is simply $\eta = x_{0}$. In what follows the notations for eigenvectors $\efield{v}_{i} = (\vpfield{v}_{i},\zeta_{i}) = (\vfield{v}_{i},\pfield{p}_{i},\zeta_{i})$ and the stable subspace component $\efield{u}_{s} = (\vpfield{u}_{s},0) = (\vfield{u}_{s},\pfield{p}_{s},0)$ are all used as appropriate. 
Substituting the solution decomposition (equation~\eqref{solution_decomposition}) into equation~\eqref{NS_CM_0_compact} one obtains
\begin{equation}
	\label{NS_CM_1}
	\begin{aligned}
	\mathsfbi{\widehat{B}}\mathsfbi{\widehat{V}}\frac{d \mathsfbi{x}}{d t} + \frac{\partial (\mathsfbi{\widehat{B}}\efield{u}_{s})}{\partial t} =&
	\mathcal{\widehat{L}}(\mathsfbi{\widehat{V}})\mathsfbi{x}  + \mathcal{\widehat{L}}(\efield{u}_{s})  
	+ \blbracket\blbracket\mathcal{N}(\mathsfbi{x},\vfield{u}_{s}) \brbracket\brbracket,
\end{aligned}
\end{equation}
where, the nonlinearities have been represented compactly inside the ceiling brackets by $\mathcal{N}(\mathsfbi{x},\vfield{u}_{s})$, defined as
\begin{equation}
	\begin{aligned}
		\mathcal{N}(\mathsfbi{x},\vfield{u}_{s}) =&
			-\left(\sum_{i=0}^{m-1}x_{i}\vecv_{i} + \vecu_{s} \right)\bcdot\nabla \left(\sum_{j=0}^{m-1}x_{j}\vecv_{j} + \vecu_{s}\right) \\
			&- x_{0}\invRec\nabla^{2} \left(\sum_{i=0}^{m-1}x_{i}\vecv_{i} + \vecu_{s} \right).
		\end{aligned} \nonumber
\end{equation}

One may multiply equation~\eqref{NS_CM_1} by $\mathsfbi{\widehat{W}}^{H}$ from the left and obtain a set of equations for the critical eigenvector amplitudes $\mathsfbi{x}$ as
\begin{eqnarray}
	\label{NS_CM_1_a}
	\frac{d \mathsfbi{x}}{d t} =&
	\mathsfbi{\widehat{\Lambda}x}  +	\mathsfbi{\widehat{W}}^{H}\blbracket\blbracket\mathcal{N}(\mathsfbi{x},\vfield{u}_{s}) \brbracket\brbracket,
\end{eqnarray}
which depend on the stable subspace solution $\vfield{u}_{s}$. Multiplying equation~\eqref{NS_CM_1_a} by $\mathsfbi{\widehat{B}}\mathsfbi{\widehat{V}}$ from the left and subtracting it from equation~\eqref{NS_CM_1} one then obtains the equation for the stable subspace solution
\begin{equation}
	\label{NS_CM_1_b}
	\frac{\partial (\mathsfbi{\widehat{B}}\efield{u}_{s})}{\partial t} =
	\mathcal{\widehat{L}}(\efield{u}_{s}) + (\mathsfbi{\widehat{I}} - \mathsfbi{\widehat{B}}\mathsfbi{\widehat{V}}\mathsfbi{\widehat{W}}^{H})\blbracket\blbracket\mathcal{N}(\mathsfbi{x},\vfield{u}_{s}) \brbracket\brbracket,
\end{equation}
where $\mathsfbi{\widehat{I}}$ is the identity matrix in the extended space. Both terms $\efield{u}_{s} = (\vpfield{u}_{s},0)$ and $\blbracket\blbracket\mathcal{N}(\mathsfbi{x},\vfield{u}_{s}) \brbracket\brbracket$ have a zero component in the extended variable therefore, the following relations hold, 
\begin{equation}
	\label{Hat_Ts_to_Ts}
	\begin{aligned}
		\mathsfbi{\widehat{B}}\efield{u}_{s} =& \blbracket \mathsfbi{B}\vpfield{u}_{s}\brbracket, \\
		\mathcal{\widehat{L}}(\efield{u}_{s}) =& \blbracket\mathcal{L}(\vpfield{u}_{s})\brbracket, \\
		(\mathsfbi{\widehat{I}} -\mathsfbi{\widehat{B}}\mathsfbi{\widehat{V}}\mathsfbi{\widehat{W}}^{H})\blbracket\blbracket\mathcal{N}(\mathsfbi{x},\vfield{u}_{s}) \brbracket\brbracket =& \blbracket(\mathsfbi{I} - \mathsfbi{B}\mathsfbi{V}\mathsfbi{W}^{H})\blbracket\mathcal{N}(\mathsfbi{x},\vfield{u}_{s}) \brbracket \brbracket.
	\end{aligned} \nonumber
\end{equation}
Therefore all terms in equation~\eqref{NS_CM_1_b} may be written without the hat notation and one pair of ceiling brackets may be dropped from the equations. A further representation change is made for the nonlinear terms such that
\begin{equation}
	\begin{aligned}
		\mathcal{\widehat{F}}(\mathsfbi{x},\vfield{u}_{s}) =& \mathsfbi{\widehat{W}}^{H}\blbracket\blbracket\mathcal{N}(\mathsfbi{x},\vfield{u}_{s})\brbracket\brbracket,  \\
		\mathcal{G}(\mathsfbi{x},\vfield{u}_{s}) =&
		\mathsfbi{Q}\blbracket\mathcal{N}(\mathsfbi{x},\vfield{u}_{s}) \brbracket, \\
		\mathsfbi{Q} =&
		(\mathsfbi{I} - \mathsfbi{B}\mathsfbi{V}\mathsfbi{W}^{H}),
	\end{aligned} \nonumber
\end{equation}
which then reduces equation~\eqref{NS_CM_0_compact} to the following set of equations,
\begin{subequations}
	\label{NS_CM_final}
	\begin{eqnarray}
		\label{CM_xk}
		\dt{\mathsfbi{x}} = 
		& \mathsfbi{\widehat{\Lambda}} \mathsfbi{x} + \mathcal{\widehat{F}}(\mathsfbi{x},\vfield{u}_{s});
	    &  \forall \ \mathsfbi{x} \in \mathds{C}^{m}, \\
		\label{CM_stable}
		\frac{\partial (\mathsfbi{B}\vpfield{u}_{s})}{\partial t} = 
		& \mathcal{L}(\vpfield{u}_{s}) +\mathcal{G}(\mathsfbi{x},\vfield{u}_{s}); 
		& \forall \ \vpfield{u}_{s} \in \mathds{T}_{s},
	\end{eqnarray}
\end{subequations}
where, $\mathds{C}^{m}$ is the complex space of dimension $m$, and, the overhead dot notation has been used to represent the time derivative, $d\mathsfbi{x}/dt$ . The result is a set of ODEs for $x_{k}$ given by equation~\eqref{CM_xk}, which represent the amplitudes of the modes lying in the critical subspace $\mathds{\widehat{T}}_{c}$, and a PDE for $\vpfield{u}_{s}$ for the evolution of the solution in the stable subspace $\mathds{T}_{s}$. Obviously $\mathds{C}^{m}$ is isomorphic to $\mathds{\widehat{T}}_{c}$. 

Equation~\eqref{NS_CM_final} is now in the appropriate sort-after form for the application of center-manifold theory. The solution to equation~\eqref{NS_CM_final} may be denoted as ($\mathsfbi{x},\vpfield{u}_{s}$) which lies in a Hilbert space $\mathds{\widetilde{H}} = \mathds{C}^{m} \bigoplus \mathds{T}_{s}$, where $\mathds{C}^{m}$ and $\mathds{T}_{s}$ represent the center and stable subspaces of the linearized operator of the reformulated system. Evolution of the critical subspace variables is given by equation~\eqref{CM_xk} and, equation~\eqref{CM_stable} represents the evolution equation for the solution field in the stable subspace. The linear operators for the two subspaces are decoupled and both the non-linear operators $\mathcal{\widehat{F}}$ and $\mathcal{G}$ contain terms that are quadratic or higher in $\mathsfbi{x}$ and $\vfield{u}_{s}$.

It is important to point out that due to the special structure of $\efield{v}^{\dagger}_{0}$ and the fact that $\lambda_{0} = 0$, the evolution equation for $x_{0}$ reduces to
	\begin{eqnarray}
			\label{CM_x0}
			\dt {x}_{0} = 0. \nonumber
\end{eqnarray}
This is not only expected but necessary since, a variation of $x_{0}$ implies a variation of the Reynolds number, which was constant by construction.

One may observe at this stage that while the reduction of the system to the final form given by equation~\eqref{NS_CM_final} required the extension of the original system and then subsequent deduction of the eigenstructure of the extended problem, the actual values of the extended variables in the eigenmodes (\ie $\zeta_{i}$ and $\zeta_{i}^{\dagger}$) play an almost trivial role. For all the direct eigenmodes $\zeta_{i}$ vanish for all $i>0$. The adjoint modes have $\zeta_{i}^{\dagger} \ne 0$, however, they are only important in projection terms of the type $\mathsfbi{\widehat{W}}^{H}\blbracket\blbracket\mathcal{N}\brbracket\brbracket$ where, the value of the extended variable vanishes for $\blbracket\blbracket\mathcal{N}\brbracket\brbracket$. Therefore $\zeta_{i}^{\dagger}$ plays no role in the projections either. Nonetheless, the consequence of considering the extended system are not trivial since the critical subspace now includes a non-trivial parameter mode $\efield{v}_{0}$. This mode does in fact play a role in $\mathcal{N}$ and  in the subsequent asymptotic evaluations of the center manifold.

\subsection{Asymptotic Approximation}

Using the Center-Manifold theorem, one may assume that $\vpfield{u}_{s}$ evolves as a graph over the critical subspace variables $\mathsfbi{x}$ such that $\vpfield{u}_{s} \sim \mathfrak{h}(\mathsfbi{x})$ where, $\mathfrak{h}$ is a function with the following properties,
\begin{eqnarray}
	\label{CM_approx_properties}
	\mathfrak{h}: \mathds{C}^{m} \to \mathds{T}_{s}, & \nonumber \\
	\mathfrak{h}(\mathsfbi{x}) \mapsto (\boldsymbol{0},\pfield{0}), & \hspace{5mm} \textit{for}\ \mathsfbi{x} = [0,0,\ldots,0] , \nonumber \\
	\mathfrak{h}(\mathsfbi{x}) \sim \mathcal{O}(\mathsfbi{x}^{2}). & \nonumber
\end{eqnarray}
\ie $\mathfrak{h}$ maps the critical subspace in to the stable subspace, vanishes at the origin and, is asymptotic to $\mathcal{O}(\mathsfbi{x}^{2})$, as $\mathsfbi{x}$ approaches the origin. 
The smoothness of $\mathfrak{h}$ depends on the smoothness of the nonlinear terms $\mathcal{\widehat{F}}$ and $\mathcal{G}$ and typically analiticity of the center-manifold can not be assumed apriori \citep{carr82,sijbrand85}. In the current work it is assumed that the nonlinear terms are smooth enough to not create any degeneracies up to the order of evaluation of the asymptotic approximations. $\mathfrak{h}$ may be substituted in to equation~\eqref{CM_stable} to obtain the graph equation,
\begin{eqnarray}
	\label{CM_graph_equation}
	\left(\frac{\partial (\mathsfbi{B}\mathfrak{h})}{\partial \mathsfbi{x}}\right)\left(\mathsfbi{\widehat{\Lambda}}\mathsfbi{x} + \mathcal{\widehat{F}}(\mathsfbi{x},\mathfrak{h}) \right) = \mathcal{L}(\mathfrak{h}) + \mathcal{G}(\mathsfbi{x},\mathfrak{h}).
\end{eqnarray}
Equation~\eqref{CM_graph_equation} must be satisfied for the graph $\mathfrak{h}$ to be an invariant center-manifold of equation~\eqref{NS_CM_final}. In this case the graph $\mathfrak{h}$ represents the field $\vpfield{u}_{s}$ that lies in the infinite-dimensional subspace $\mathds{T}_{s}$. The matrix $\mathsfbi{B}$ in the derivative arises due to the special nature of the incompressibility equation.

Equation~\eqref{CM_graph_equation} is in general a hard problem to solve. As an alternative one may approximate $\mathfrak{h}$ asymptotically via a power series in $\mathsfbi{x}$ as
\begin{alignat}{4}
	\label{CM_H_approx}
		\mathfrak{h}(\mathsfbi{x}) = \sum_{a=0}^{m-1}\sum_{b=a}^{m-1} x_{a}x_{b}\vpfield{y}_{a,b} + 
		 \mathcal{O}(\mathsfbi{x}^{3}),
\end{alignat}
where the various $\vecyp$'s represent the (\emph{time independent}) velocity and pressure fields associated with the corresponding coefficients of $\mathsfbi{x}$ in the power series, \ie\ $\vecyp_{0,0} = (\vfield{y},\pfield{p})_{0,0}$ \textit{etc}. The power series for $\mathfrak{h}$ may be substituted into equation~\eqref{CM_graph_equation} and the expressions for the various powers of the terms may be collected to obtain an equation governing each field $\vecyp$.

Before writing the an explicit expressions for the solutions a key feature of the approximation is highlighted with an example. Suppose one is only interested in a second-order approximation of $\mathfrak{h}$. We know that $\mathfrak{h} \sim \mathcal{O}(\mathsfbi{x}^{2})$ which implies $\partial \mathfrak{h}/\partial \mathsfbi{x} \sim \mathcal{O}(\mathsfbi{x})$. Therefore the first term on the LHS $(\partial \mathfrak{h}/\partial \mathsfbi{x})\mathsfbi{\widehat{\Lambda}}\mathsfbi{x} \sim \mathcal{O}(\mathsfbi{x}^{2})$ and will have contributions at second order. The contributing terms however, are all linear in $\mathbb{y}$. The second term on the LHS is at least of $\mathcal{O}(\mathsfbi{x}^{3})$, \ie $(\partial \mathfrak{h}/\partial \mathsfbi{x})\mathcal{\widehat{F}}(\mathsfbi{x},\mathfrak{h}) \sim \mathcal{O}(\mathsfbi{x}^{3})$, since $\mathcal{\widehat{F}}(x,\mathfrak{h})$ is atleast quadratic in $\mathsfbi{x}$. Hence it does not contribute at second order. On the RHS, the first term $\mathcal{L}(\mathfrak{h}) \sim \mathcal{O}(\mathsfbi{x}^{2})$ since $\mathcal{L}$ only operates on $\mathbb{y}$ and does not change the order of $\mathsfbi{x}$. Again, the resulting expressions are all linear in $\mathbb{y}$. The second term on the RHS needs a bit more consideration. Since $\mathcal{G}(\mathsfbi{x},\mathfrak{h})$ was originally quadratic, it will have three types of expressions -- quadratic terms of the type $\mathsfbi{x}\cdot\mathsfbi{x}$, quadratic terms of the type $\mathfrak{h}\cdot\mathfrak{h}$, and mixed quadratic terms of the type $\mathsfbi{x}\cdot\mathfrak{h}$. Clearly, $\mathfrak{h}\cdot\mathfrak{h} \sim \mathcal{O}(\mathsfbi{x}^{4})$ and $\mathsfbi{x}\cdot\mathfrak{h} \sim \mathcal{O}(\mathbb{x}^{3})$. Thus the only second order terms that arise are due to the terms originally of the type $\mathsfbi{x}\cdot\mathsfbi{x}$. These terms however are independent of $\mathbb{y}$, since they multiply the eigenvectors of the critical subspace. Thus at second order we have no non-linearities in $\mathbb{y}$ and all equations for the unknown fields $\mathbb{y}$ are inhomogeneous linear equations. This feature is not restricted to second order. At all orders the resulting equations for $\mathbb{y}$ are inhomogeneous linear equations and thus the highly nonlinear problem for $\mathfrak{h}$ is reduced to a series of linear problems for $\mathbb{y}$ when approximating as a power series in $\mathsfbi{x}$. This is perhaps not immediately obvious at the onset however, it is a general feature of asymptotic power series approximations and is perfectly analogous to the case of asymptotic center-manifold approximations for systems described by ODEs, where one obtains a series of linear equations for the coefficients of the various powers of the critical variables. With a slight abuse of terminology, the various fields $\vpfield{y}$ may indeed be considered the infinite-dimensional ``coefficients'' of the terms in power series in $\mathsfbi{x}$. 

After substitution of the asymptotic approximation~\eqref{CM_H_approx} into the graph equation~\eqref{CM_graph_equation} and performing the algebraic manipulation to collect similar terms together, one obtains the following expressions for the second order approximations
\begin{equation}
	\label{second_order_expressions}
	\begin{aligned}
		[(\lambda_{i} + \lambda_{j})\mathsfbi{B} -  \mathcal{L}]\vpfield{y}_{i,j} = &
		-\mathsfbi{Q}\blbracket (\invRec\nabla^{2}\vfield{v}_{j})\delta_{i,0}\brbracket \\
		 & +\mathsfbi{Q}\blbracket (\vfield{v}_{i}\bcdot\nabla\vfield{v}_{j}) \delta_{i,j}\brbracket \\
		& -\mathsfbi{Q}\blbracket\vfield{v}_{i}\bcdot\nabla\vfield{v}_{j}
		+\vfield{v}_{j}\bcdot\nabla\vfield{v}_{i}
		\brbracket,
	\end{aligned}  
\end{equation}
for $i,j \in 0\ldots (m-1), \  j\ge i$, where, $\delta_{i,j}$ is the Kronecker-Delta function. Recalling that $\vpfield{y}\in\mathds{T}_{s}$, it is appropriate to include a projection operator $\mathsfbi{P}^{\obliquesymbol{}} = (\mathsfbi{I} - \mathsfbi{V}\mathsfbi{W}^{H}\mathsfbi{B})$ in the final expressions to ensure that $\vpfield{y}$ lies in the prescribed invariant subspace. The inherent structure of the problem is obvious. If the second-order inhomogeneous term is generically represented as $\vpfield{f}_{i,j}$, one may represent the various solutions as
\begin{eqnarray}
	\label{general_second_order_expression}
	\begin{aligned}
		\vpfield{y}_{i,j} &= \mathcal{R}_{s}(\lambda_{i} + \lambda_{j})\vpfield{f}_{i,j}; &i,j \in 0\ldots (m-1), \  j\ge i, && \\
		\mathcal{R}_{s}(\omega) &= \mathsfbi{P}^{\obliquesymbol{}}[\omega\mathsfbi{B} -  \mathcal{L}]^{-1}. &  &&
	\end{aligned}
\end{eqnarray}
Here $\mathcal{R}_{s}(\omega)$ may be interpreted as the restricted resolvent operator. The structure persists at higher orders of approximation so that the third order approximation may be written as 
\begin{equation}
	\label{general_higher_order_expression}
	\begin{aligned}
		\vpfield{y}_{i,j,k} &= \mathcal{R}_{s}(\omega_{i,j,k})\vpfield{f}_{i,j,k}; 
		& \omega_{i,j,k} = \lambda_{i} + \lambda_{j} + \lambda_{k}, 		&&
	\end{aligned}
\end{equation}
and so on, where $\vpfield{f}$ includes contributions from solution fields $\vpfield{y}$ evaluated at lower orders. Obviously this generalizes to any finite dimension $m$ of the critical subspace. However, the number of terms at each order rises rapidly as $m$ increases. The inhomogeneous term $\vpfield{f}$ depends on the structure of the nonlinearity and becomes increasingly cumbersome at higher orders of approximations. It is likely that a general structure may be found for the non-linear terms in $\vpfield{f}$ as well, however, that has not been attempted in the current work. 

In general the resolvent operator of a linear system has poles in the complex plane at the points corresponding to the eigenvalues of the linear system. In this case, the linear operator in question is the restriction of the original operator to the stable subspace. Therefore all the poles of the restricted resolvent lie in the left half of the complex plane, excluding the imaginary axis. Or more generally, no poles of the restricted resolvent lie along the imaginary axis. On the other hand, the angular frequency $\omega$ in equations~\eqref{general_second_order_expression} and \eqref{general_higher_order_expression} always lies on the imaginary axis since it is comprised of summations of the integer multiples of the critical angular frequencies. Hence equations~\eqref{general_second_order_expression} and \eqref{general_higher_order_expression} are never singular. The possibility of singularity of the full resolvent operator is associated with the appearance of secular terms when dealing with multiple time-scale expansions \citep{bender99}, or with the generation of higher-order terms when evaluating normal forms \citep{wiggins03,guckenheimer83,coullet83,haragus11,carini15}. In the current case, the question of resonance never arises.

Once $\mathfrak{h}$ is evaluated to the desired order, it may be substituted back in to the expression for $\mathcal{\widehat{F}}$ in equation~\eqref{CM_xk} and the result is a set of ``amplitude equations'' for the extended system. Note that these are slightly different from the Stuart-Landau equations which usually refer to the slowly evolving part of the mode in question, derived usually through an assumption of scale separation \citep{newell69,cross09,sipp07}. Here no such separation of time scales has been assumed and the amplitudes include the ``fast'' oscillatory behavior as well as any slow modulation that may emerge. To distinguish, the equations will be refered to as the center-manifold amplitude equations, with the variables $x_{i}$ naturally corresponding to the amplitudes of the modes in the center-manifold. 

$\mathcal{\widehat{F}}$ may be written as individual components, $\mathcal{\widehat{F}}_{k} = \langle\efield{v}^{\dagger}_{k},\blbracket\blbracket\mathcal{N}(\mathsfbi{x},\mathfrak{h})\brbracket\brbracket\rangle$. Recall that that $\dt{x}_{0} = 0$, so only $k\in1,\ldots,(m-1)$ is relevant. In the subsequent expressions the ceiling brackets are dropped due to the relation, $\langle\efield{v}^{\dagger}_{k},\blbracket\blbracket\mathcal{N}\brbracket\brbracket\rangle = \langle\vfield{v}^{\dagger}_{k},\mathcal{N}\rangle$, where $\vfield{v}^{\dagger}_{k}$ is now the velocity field of the $k^{th}$ adjoint eigenvector of the \emph{standard} problem. Assuming a second-order approximation for $\mathfrak{h}$ has been evaluated, this results in the following expression for the center-manifold equations
\begin{equation}
	\label{extended_amplitude_equations}
	\begin{aligned}
		\dt{x}_{k} =& \lambda_{k}x_{k}  + \mathcal{\widehat{F}}_{k}^{(2)} + \mathcal{\widehat{F}}_{k}^{(3)} + \mathcal{O}(\mathsfbi{x^{4}}),
	\end{aligned}
\end{equation}
for $k\in 1,\ldots,(m-1)$. Here, $\mathcal{\widehat{F}}_{k}^{(2)}$ and $\mathcal{\widehat{F}}_{k}^{(3)}$ are the second and third order nonlinear terms respectively for the $k^{th}$ center-manifold amplitude equation, defined as,
\begin{subequations}
	\label{amplitude_second_oder}
	\begin{eqnarray}
		\label{amplitude_f2}
		\mathcal{\widehat{F}}_{k}^{(2)} = &&\left\lbrace
		\begin{aligned}
		& -\sum_{i=0}^{m-1}\sum_{j=i}^{m-1} x_{i}x_{j}\invRec\langle\vfield{v}^{\dagger}_{k}, \nabla^{2}\vfield{v}_{j}\rangle\delta_{i,0} \\
		& - \sum_{i=0}^{m-1}\sum_{j=0}^{m-1} x_{i}x_{j}\langle\vfield{v}^{\dagger}_{k}, \vfield{v}_{i}\bcdot\nabla\vfield{v}_{j}\rangle,
	\end{aligned} \right. \\
	\label{amplitude_f3}
	\mathcal{\widehat{F}}_{k}^{(3)} = &&\left\lbrace
	\begin{aligned}
			& -\sum_{i=0}^{m-1}\sum_{a=0}^{m-1}\sum_{b=a}^{m-1} x_{i}x_{a}x_{b}\invRec\langle\vfield{v}^{\dagger}_{k}, \nabla^{2}\vfield{y}_{a,b}\rangle\delta_{i,0} \\
			& -\sum_{i=0}^{m-1}\sum_{a=0}^{m-1}\sum_{b=a}^{m-1} x_{i}x_{a}x_{b}\langle\vfield{v}^{\dagger}_{k}, \vfield{v}_{i}\bcdot\nabla\vfield{y}_{a,b}\rangle \\
			& -\sum_{i=0}^{m-1}\sum_{a=0}^{m-1}\sum_{b=a}^{m-1} x_{a}x_{b}x_{i}\langle\vfield{v}^{\dagger}_{k}, \vfield{y}_{a,b}\bcdot\nabla\vfield{v}_{i}\rangle.
		\end{aligned} \right. 
	\end{eqnarray}	
\end{subequations}
Equation~\eqref{extended_amplitude_equations} contains terms of the type $(x_{0})^{p}, \ p\ge2$. Since $x_{0}$ represents the parameter variation these terms are constant once the parameter value has been chosen and in effect act as inhomogeneous terms for the center-manifold equations. 

It is important to highlight the key aspect of the procedure which considers the extended problem for the application of the center-manifold theorem. If one proceeds with the standard problem without the extended critical subspace, then equation~\eqref{CM_graph_equation} is still valid for the standard problem, but only at the bifurcation point. At this stage, the complexity of the problem for the solution of the graph equation is not significantly different whether one considers the standard or the extended problem to derive the graph equation. The graph equation is for the solution lying in the stable subspace and, as has been pointed out, the subspace expansion only affects the critical subspace and not the stable subspace. Therefore $\mathfrak{h}$ needs to be approximated formally as an asymptotic expansion regardless. The difference arises when one considers problems that are perturbed slightly away from the bifurcation point. Now the asymptotic solutions obtained for the standard problem are no longer valid and one must consider a second series expansion (in the bifurcation parameter) to evaluate the continuation of the previously calculated asymptotic terms. In addition to the added complexity, this has the unaesthetic quality of considering a series on top of an already trucated series. Regardless of one's aesthetic inclination, this added complexity is clearly avoided when one considers the extended problem which already includes the effects of parameter variation. An alternate way around the problem of double series expansion has been used by \cite{haragus11} wherin one starts with the expected normal form of the reduced problem, deduced from the spectrum of the problem at bifurcation and then evaluates the asymptotic terms satisfying the normal form. This can be done for simple bifurcation problems however, for more complicated bifurcation problems in higher codimensions one may not be able to assume the normal form apriori. Additionally, more often than not the whole point of such analysis is to deduce the type of dynamics that are induced by the bifurcation, and not start with an apriori assumption of the dynamics. 

Hence, the method derived here with the extended problem shines when considering center-manifold problems that have been perturbed away from the bifurcation point. It can be considered as a way of incorporating the double asymptotic series expansion (which is formally necessary when considering the standard problem) into a single asymptotic expansion, without an apriori assumption of the normal form of the reduced dynamics. The cost of unifying the two asymptotic expansions is to consider a larger critical subspace.

%% file: cyl.tex
\section{Hopf bifurcation in a cylinder wake}
\label{application_cylinder}

\subsection{The Extended Problem}

The asympotic formulation derived in the previous sections is applied to the case of the first bifurcation of a cylinder wake. The bifurcation problem has been investigated by several authors in various settings \citep{provansal87,dusek94,pier02,barkley06,giannetti07,sipp07,mantivc15,negi20}. The problem is revisited in the context of a center manifold approximation which is valid in the vicinity of the bifurcation point. 

The investigation is carried out numerically using Nek5000, an open source fluid dynamics solver originating at Argonne National Laboratory \citep{nek5000}. The code implements the spectral element method \citep{patera84}  for the Navier--Stokes equations with the $\mathds{P}_{n}-\mathds{P}_{n-2}$ formulation \citep{maday89}. The current work uses ninth order Lagrange polynomials in each direction for the representation of velocities in each spectral element and a seventh order Lagrange polynomial for the pressure with a total of $3760$ elements in the whole domain. Time integration of the system is performed using a third-order backward difference scheme with an implicit treatment of the viscous term. The non-linear terms are evaluated explicitly using extrapolation. Incompressibility is enforced using the pressure projection method \citep{deville02} implemented through the inexact LU factorization. Numerical stabilization is done using the HPF-RT method \citep{negiphd}, which adds a small amount of dissipation at the highest wavenumber scales. All quantities are non-dimensionalized using the fluid density $\rho$, the cylinder diameter $D$ and the inflow velocity $U$. The flow is characterized by the Reynolds number defined as $\Rey = UD/\nu$, where $\nu$ is the kinematic viscosity of the fluid. 

The problem domain is set up with the cylinder center located at the origin, the inflow boundary located at $x=-20$, the outflow boundary located at $x=40$ and far field boundaries located at $y=\pm20$. Dirichlet boundary condition is specified at the inflow and the standard no-stress boundary condition is applied at the outflow. The far-field boundary is equipped with the symmetry boundary condition. The bifurcation point of the system is found via a combination of selective fequency damping \citep{akervik06} for determining the fixed point of the system at each $\Rey$, along with a bisection-like algorithm for determining the point with zero growth rate. The spectral problem was solved using the Krylov-Schur algorithm \citep{stewart02}. The details of the procedure may be found in Appendix B.

\begin{figure}
	\centering
	\includegraphics[width=0.49\textwidth]{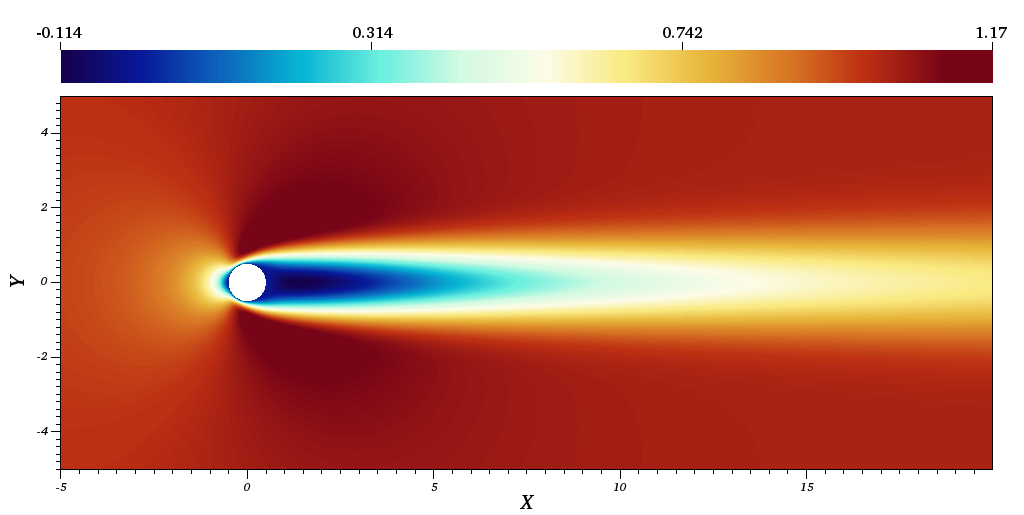}
	\includegraphics[width=0.49\textwidth]{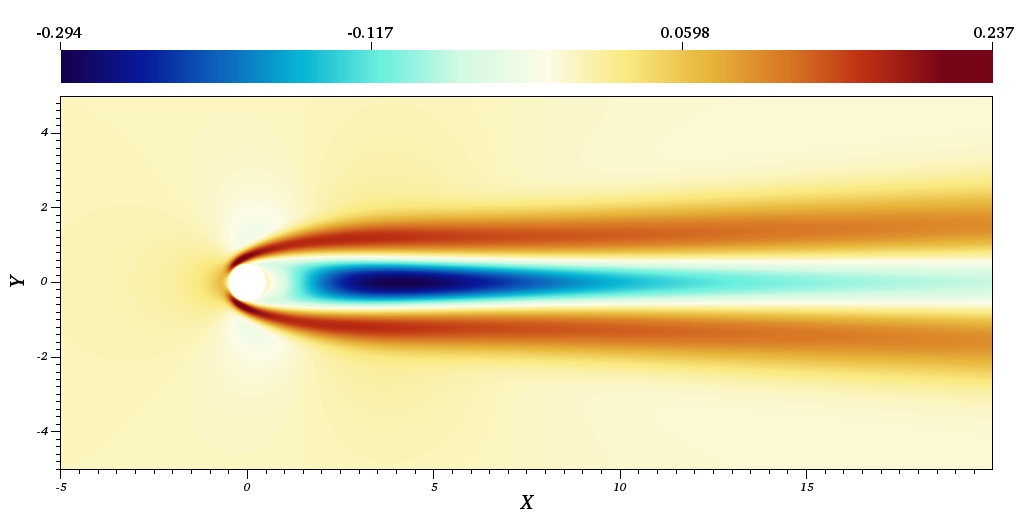}
	\caption{Streamwise velocity of (top) the stationary base flow at $\Rey_{c}=46.30$ and (bottom) the parameter mode.}
	\label{fig:base_param}
\end{figure}

\begin{figure}
	\centering
	\includegraphics[width=0.49\textwidth]{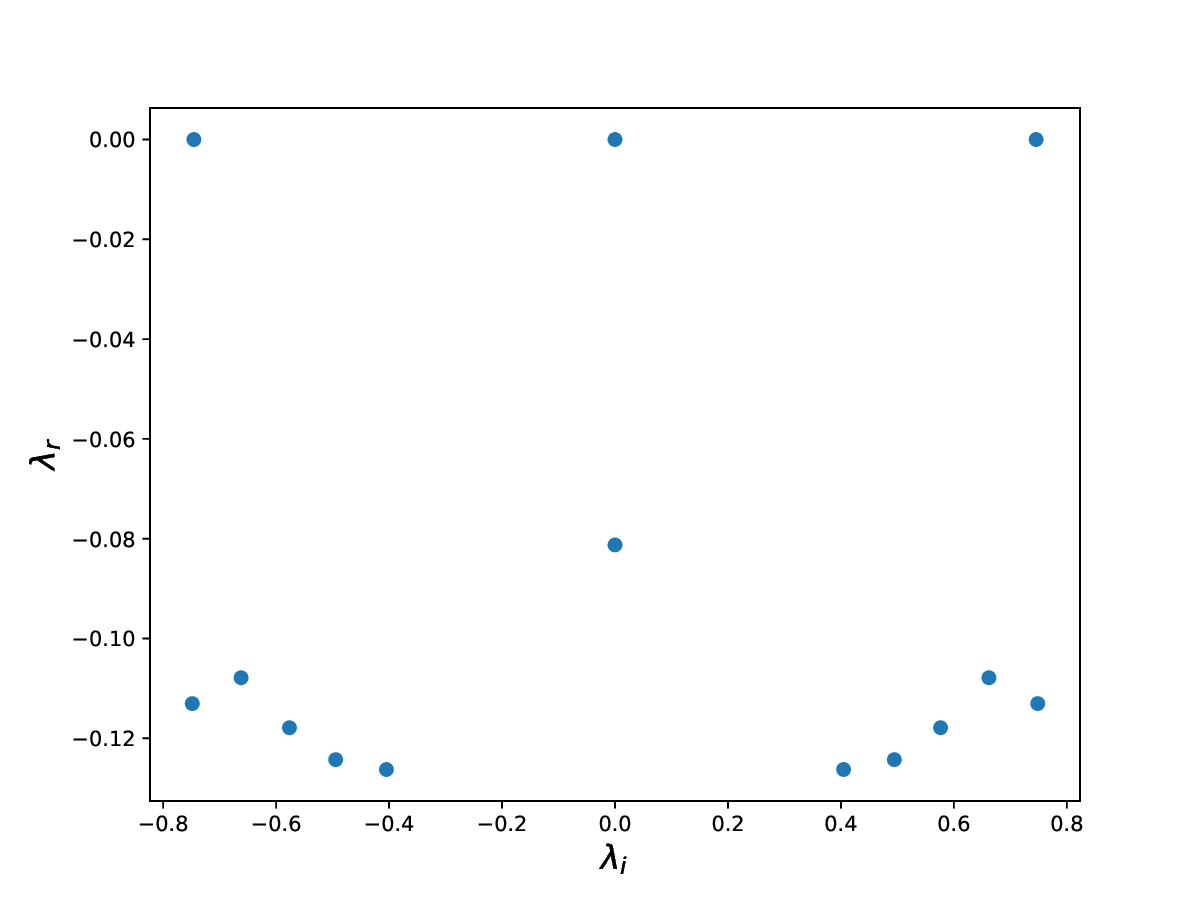}
	\caption{Spectrum for the flow across a cylinder at the critical Reynolds number of $Re_{c}=46.30$. The spectrum includes the parameter mode located at the origin.}
	\label{fig:cyl_spectrum}
\end{figure}

\begin{figure}
	\includegraphics[width=0.49\textwidth]{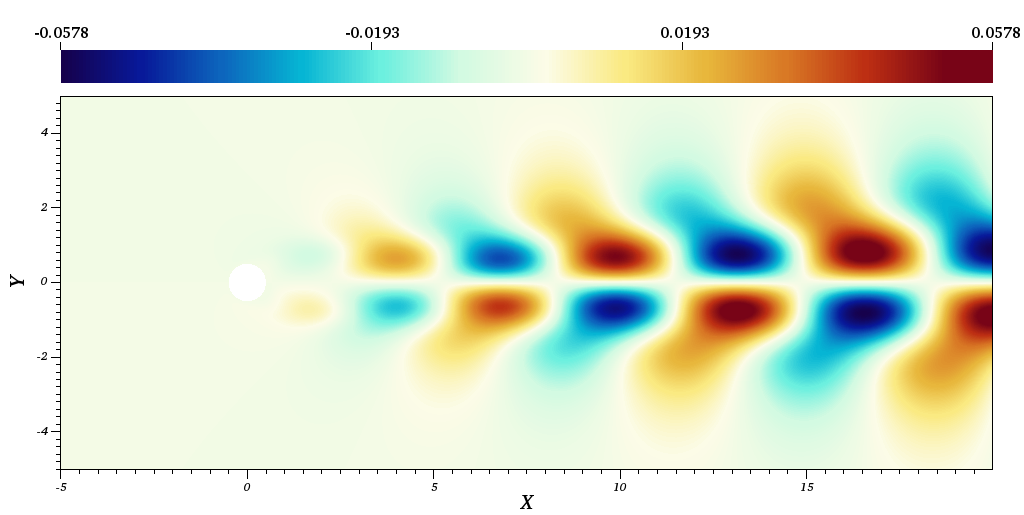}
	\includegraphics[width=0.49\textwidth]{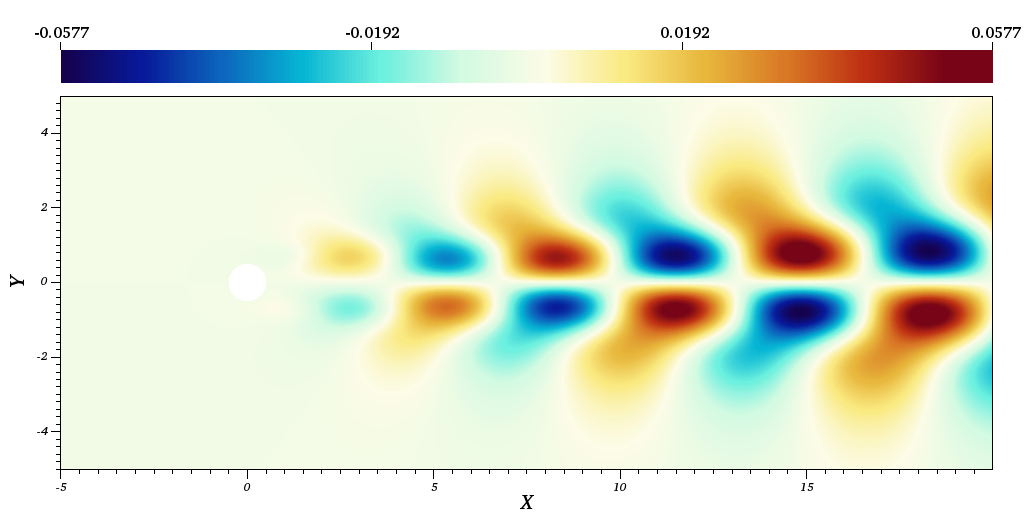}
	\caption{Streamwise velocity of the direct eigenmode. The panels represent the (top) real part and (bottom) imaginary part of the direct critical mode.}
	\label{fig:flowconfig}
\end{figure}

\begin{figure}
	\includegraphics[width=0.49\textwidth]{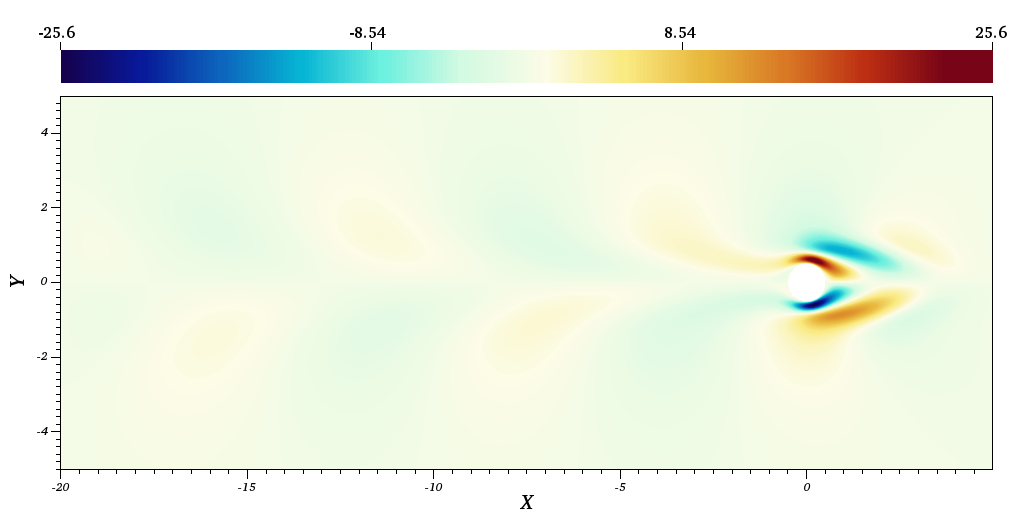}
	\includegraphics[width=0.49\textwidth]{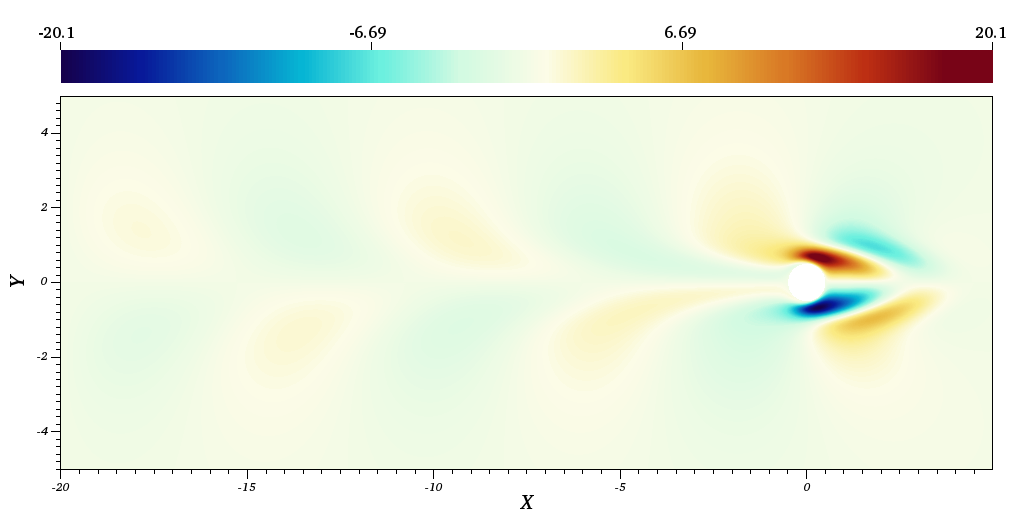}
	\caption{Streamwise velocity of the adjoint eigenmode. The panels represent the (top) real part and (bottom) imaginary part of the adjoint critical mode.}
	\label{fig:flowconfig_2}
\end{figure}

The critical Reynolds number is found to be $\Rey_{c} = 46.30$ which is close to the value of $46.6$ reported in the work of \cite{sipp07}. The spectrum obtained at the bifurcation point is depicted in figure~\ref{fig:cyl_spectrum}, including the parameter mode located at the origin. The streamwise velocity for the calculated baseflow along with the parameter mode due to the extension of the system is depicted in figure~\ref{fig:base_param}. The complex conjugate pair of critical eigenvalues for the baseflow is found to be $\lambda_{c} = \pm0.7456i$, which compares well with $\lambda_{c}=\pm0.74i$ reported in \cite{sipp07}. The direct and adjoint critical eigenmodes of the flow are shown in figures~\ref{fig:flowconfig} and ~\ref{fig:flowconfig_2} respectively. The complex-conjugated mode is not shown. The modes are normalized such that the direct modes are of unit norm and the adjoint modes are scaled appropriately to maintain biorthogonality. The parameter mode is the exception where the normalization is such that $\zeta^{\dagger}_{0} = \zeta_{0} = 1$.

\subsection{Asymptotic Center Manifold Approximation}

\begin{figure}
	\centering
	\includegraphics[width=0.49\textwidth]{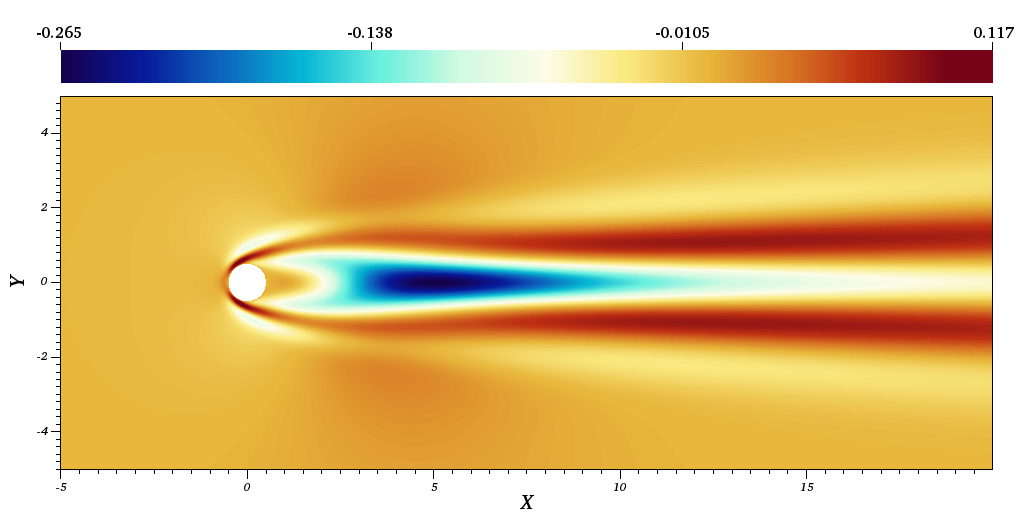}
	\includegraphics[width=0.49\textwidth]{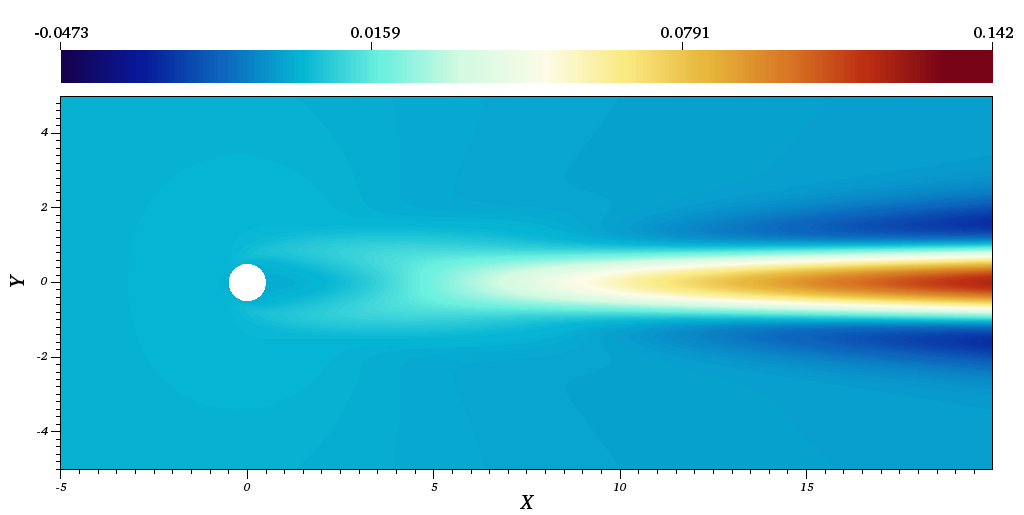}							
	\caption{Streamwise velocity components of the restricted resolvent solutions corresponding to the zero frequency fields (top) $\vpfield{y}_{0,0}$ and (bottom) $\vpfield{y}_{1,2}$.}
	\label{fig:resolvents_zero}
\end{figure}

For a second order asymptotic approximation of the graph $\mathfrak{h}(\mathsfbi{x})$, one obtains six fields $\vpfield{y}_{i,j}$, that are solutions to the restricted resolvent equations~\eqref{second_order_expressions}, corresponding to the fields associated with the terms $x_{i}x_{j}$ in the series expansion of $\mathfrak{h}$, where, $i,j\in 0,1,2;\ j \ge i $. The fields are in general complex with the exception of $\vpfield{y}_{0,0}$ and $\vpfield{y}_{1,2}$, which are restricted resolvent solutions at zero frequency and are purely real fields. The fields $\vpfield{y}_{0,0}$ and $\vpfield{y}_{1,2}$ are depicted in figure~\ref{fig:resolvents_zero}. These, in combination with the parameter mode, provide contributions to what is often termed as the mean flow correction.

\begin{figure}
	\centering
	\includegraphics[width=0.49\textwidth]{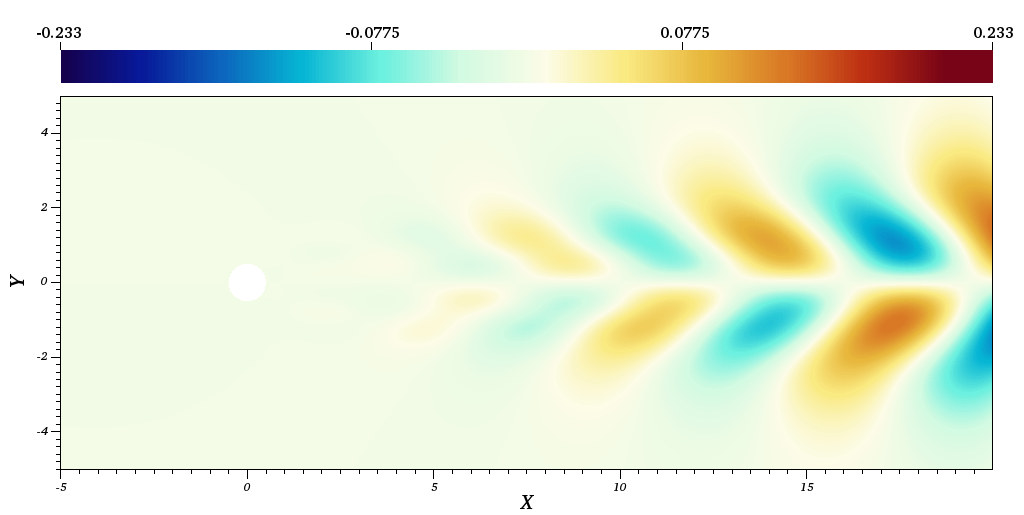}
	\includegraphics[width=0.49\textwidth]{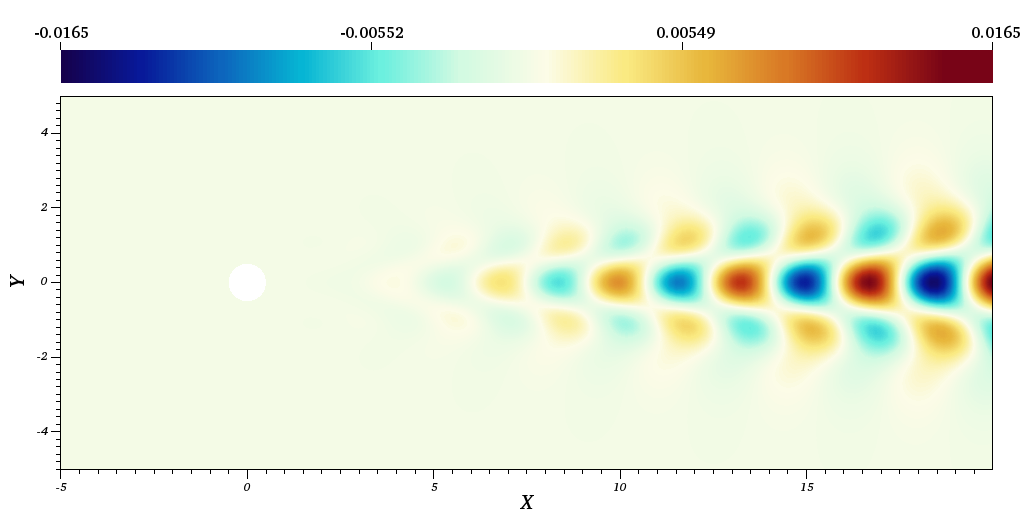}
	\caption{The panels depict the real part of the streamwise velocity components of (top) $\vpfield{y}_{0,1}$ and (bottom) $\vpfield{y}_{1,1}$ with angular frequencies equal to the imaginary parts of $\lambda_{c}$ and $2\lambda_{c}$.}
	\label{fig:resolvents_oscillatory_real}
\end{figure}

\begin{figure}
	\centering
	\includegraphics[width=0.49\textwidth]{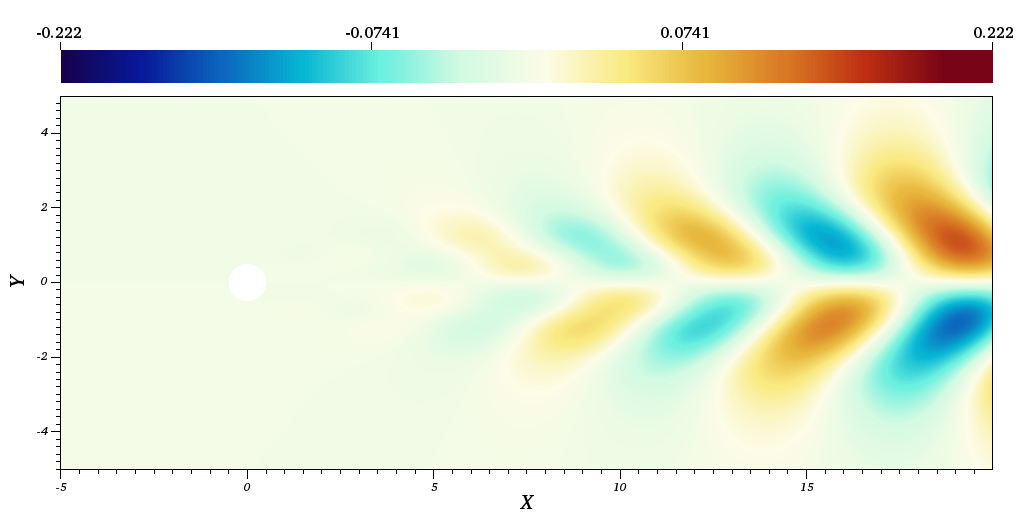}
	\includegraphics[width=0.49\textwidth]{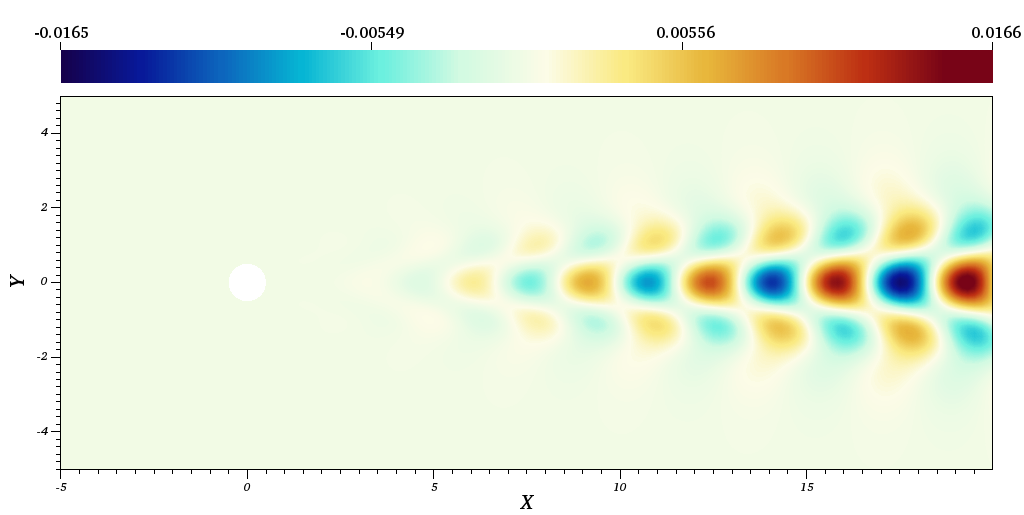}
	\caption{The panels depict the imaginary part of the streamwise velocity components of (top) $\vpfield{y}_{0,1}$ and (bottom) $\vpfield{y}_{1,1}$ with angular frequencies equal to the imaginary parts of $\lambda_{c}$ and $2\lambda_{c}$.}
	\label{fig:resolvents_oscillatory_imag}
\end{figure}

\begin{figure}
	\centering
	\includegraphics[width=0.5\textwidth]{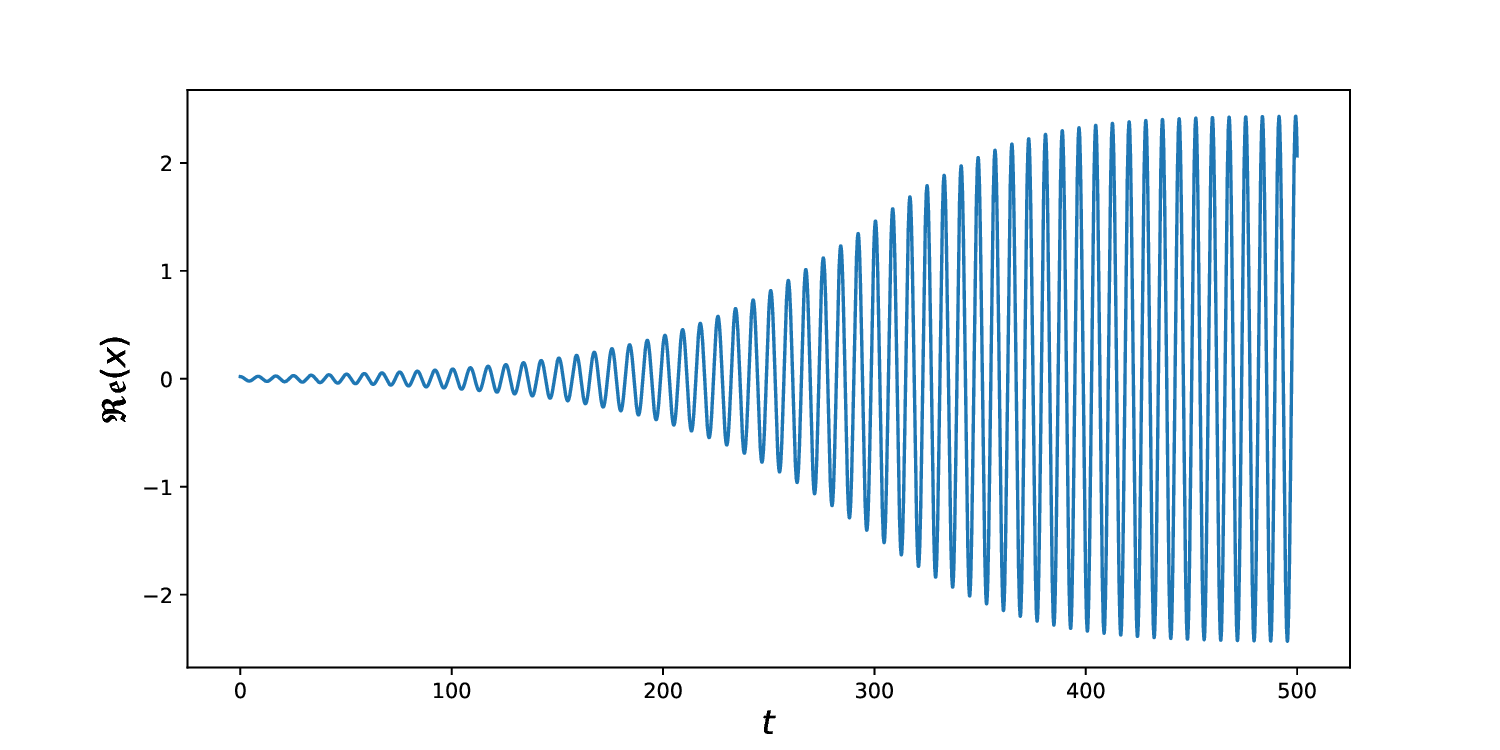}	
	\caption{The time evolution of the real part of $x$ via the center-manifold amplitude equations. The system is evaluated at $\eta=0.074$}
	\label{fig:center_manifold_evolution}
\end{figure}

\begin{figure}
	\centering
	\begin{subfigure}[t]{0.49\textwidth}
		\includegraphics[width=1.0\textwidth]{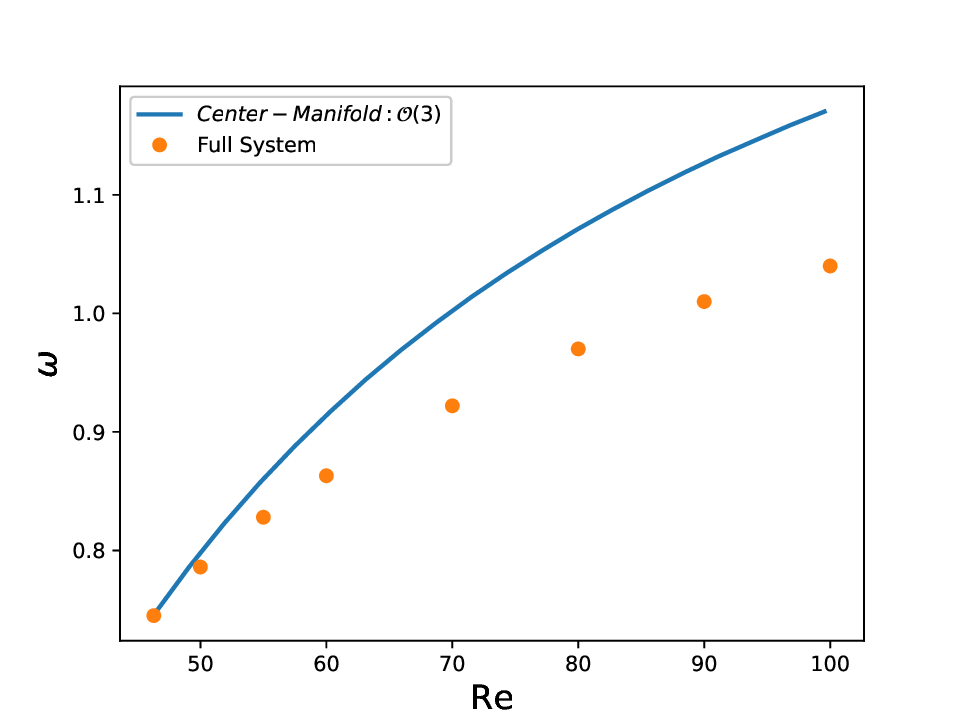}
		\caption{}
		\label{fig:limit_cycle_frequency}
	\end{subfigure}	
	\begin{subfigure}[t]{0.49\textwidth}
		\includegraphics[width=1.0\textwidth]{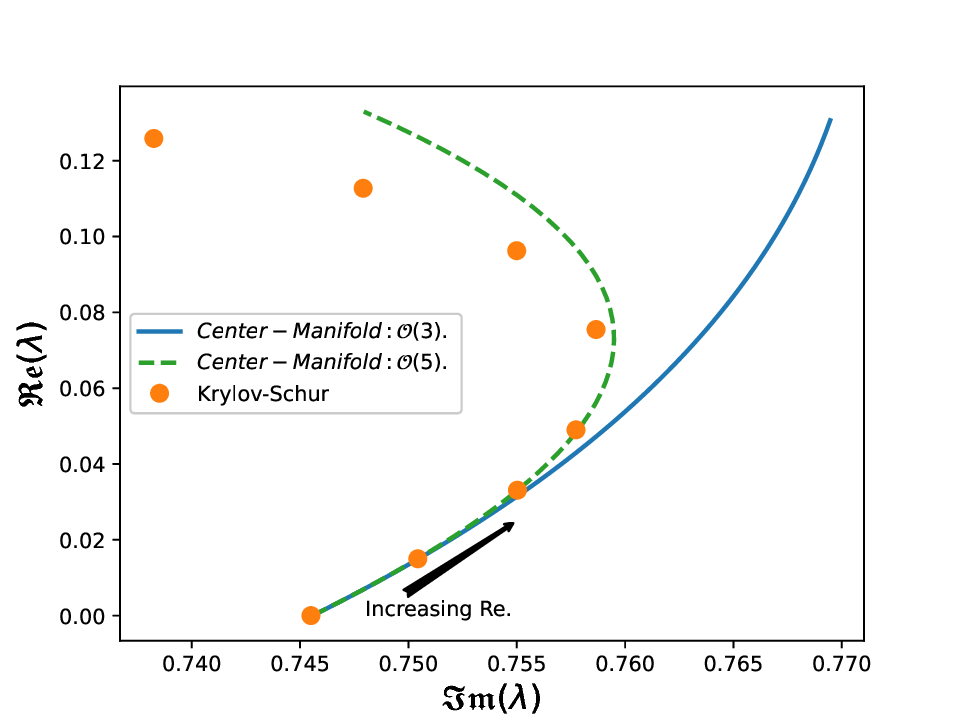}
		\caption{}
		\label{fig:linearized_eigenvalue}
	\end{subfigure}									
	\caption{Comparison of the linear and non-linear angular frequencies of the full system and the center-manifold amplitude equations. (a) Comparison of the angular frequencies of the saturated limit cycles at different Reynolds numbers. (b) Comparison of the linearized system eigenvalue (with positive imaginary part) for varying Reynolds numbers.}
	\label{fig:frequency_comparison}
\end{figure}

The real and imaginary parts of the resolvent solution fields $\vpfield{y}_{0,1}$ and $\vpfield{y}_{1,1}$ are depicted in figure~\ref{fig:resolvents_oscillatory_real} and ~\ref{fig:resolvents_oscillatory_imag}. The fields $\vpfield{y}_{0,1}$ and $\vpfield{y}_{0,2}$ are complex-conjugate pairs, and likewise for $\vpfield{y}_{1,1}$ and $\vpfield{y}_{2,2}$ (the complex conjugated fields are not displayed). These are the oscillatory solutions of the second order asymptotic approximation of $\mathfrak{h}$. The angular frequencies of $\vpfield{y}_{0,1}, \vpfield{y}_{0,2}, \vpfield{y}_{1,1}, \vpfield{y}_{2,2}$ are the imaginary parts of $\lambda_{c},-\lambda_{c},2\lambda_{c}$ and $-2\lambda_{c}$ respectively. 
Once the fields $\vpfield{y}_{i,j}$ have been evaluated, substitution in to equation~\eqref{amplitude_second_oder} then results in the coefficients of the polynomial terms $x_{i}x_{j}$ and $x_{i}x_{j}x_{k}, \ (i,j,k \in 0,1,2;\ j\ge i; k\ge j)$ of the center-manifold amplitude equations. When the graph equation is evaluated asymptotically at second order the resulting center-manifold equations are of third order (in $\mathsfbi{x}$). Since $x_{1}$ and $x_{2}$ are the amplitudes of the complex-conjugated pair of modes, the resulting expressions for the center-manifold amplitude equations obtained for the two variables are also complex-conjuated with respect to the other. Differences arise due to the numerical accuracy of the terms evaluated and in the current work this complex-conjugacy property is violated at $\mathcal{O}(10^{-8})$. A full list of the evaluated terms is given in the appendix. The combinatorial problem would allow a total of $19$ terms for each equation however, a significant number of the terms are of the order of the convergence tolerance used for the spectral solver and are treated as negligible. Even the inhomogeneous terms $x_{0}x_{0}$ and $x_{0}x_{0}x_{0}$ are negligible in the current case under examination. The vanishing of several terms in the center-manifold amplitude equations may also be deduced by inspecting the spatial symmetry of the resulting terms. Replacing $x_{1},x_{2}$ with $x,x^{*}$, and $x_{0}$ with the parameter variable $\eta$, and ignoring the small terms ($<10^{-5}$), one obtains the following equation for $\dt{x}$
\begin{equation}
	\label{numerical_amplitude_eqn_x1_p1}
	\begin{aligned}
			\dt{x} =& +[0.7456i]x  \\
			 &   + [(1.976 + 0.698i)\times10^{-1}]\eta x \\ 
			& + [(6.388 + 4.616i)\times10^{-3}]\eta x^{*}  \\
			 & + [(7.264 - 4.787i)\times10^{-2}]\eta^{2} x \\
			& + [(4.477 - 1.482i)\times10^{-3}]\eta^{2}x^{*} \\
			 & + [(1.615 - 2.904i)\times10^{-5}]x^{3}  \\
			& - [(2.526 - 8.010i)\times10^{-3}](x^{*}x)x \\
			 & + [(0.973 - 3.380i)\times10^{-5}](x^{*}x)x^{*} \\
		\end{aligned}
\end{equation}
The complex conjugation of equation~\eqref{numerical_amplitude_eqn_x1_p1} results in the equation for $\dt{x}^{*}$. The equation may now be evaluated numerically to obtain the time history and the resulting non-linear angular frequency of the reduced system. The evolution of the system at a parameter value of $\eta=0.074$ ($\Rey=50.0$) is shown in figure~\ref{fig:center_manifold_evolution}, depicting the typical evolution to the saturated limit cycle. 
The center-manifold amplitude equations are numerically integrated for several different parameter values and the angular frequency of the limit cycle oscillations is obtained. The comparison between the frequencies obtained through the full non-linear simulations and those obtained through the reduced center-manifold amplitude equations is show in figure~\ref{fig:limit_cycle_frequency}. Close to the bifurcation point the saturated limit-cycle frequencies are well predicted by the reduced equations however, as the Reynolds number increases the saturation frequency systematically deviates from the results of the full non-linear simulations. Recall that the the center-manifold reduction is only an asymptotic approximation, therefore higher order terms could provide contributions as the system is moved away from the critical point. Alternately, the center-manifold approximation itself may not be valid sufficiently far away from the critical point, so that the solution $\vpfield{u}_{s}$ which lies in the stable subspace can no longer be assumed to evolve as a graph of the critical subspace. 

In addition to the saturated limit cycle angular frequency one may also estimate the variation of the linearized angular frequency \ie the eigenvalue of system beyond the critical point through the reduced equations. Since the inhomogeneous terms in the reduced system are negligible, the variation of the eigenvalue as a function of the parameter $\eta$ is directly obtained from the linear terms  in equation~\eqref{numerical_amplitude_eqn_x1_p1}. A comparison of the obtained eigenfrequencies of the center-manifold equations ($\mathcal{O}(\mathsfbi{x}^{3})$) and those computed for the full system using the Krylov-Schur algorithm for different Reynolds numbers is shown in figure~\ref{fig:linearized_eigenvalue}. Close to the critical point the variation of the eigenvalue seems to be fairly well captured by the reduced equations. In an attempt to obtain a better asymptotic approximation higher order terms up to $\mathcal{O}(\mathsfbi{x}^{4})$ were evaluated resulting in a fifth order equation for the center-manifold amplitudes. However, at this order the highest order term $(|x|^{4}x)$ has a coefficient with a positive real part, leading to an unbounded system. Presumably one needs to evaluate $\mathfrak{h}(\mathsfbi{x})$ to $\mathcal{O}(\mathsfbi{x}^6)$ to obtain a bounded reduced system again (in which case the center-manifold amplitude equations would be of $\mathcal{O}(\mathsfbi{x}^{7})$). Further higher order approximations have not been pursued. One may nonetheless obtain a higher-order estimate of the linearized frequencies of the reduced system from this higher order (but unbounded) system ($\mathcal{O}(\mathsfbi{x}^{5})$). This is plotted in figure~\ref{fig:linearized_eigenvalue} with a dashed line. Clearly the higher order estimate follows the eigenvalues of the full system much more closely, providing evidence for the conjecture that the higher order terms of the asymptotic approximation do indeed have significant contribution as the system moves away from the critical point.

\subsection{The Stuart-Landau equation}
\label{stuart_landau}

The Landau model has often been utilized to understand the behavior of weakly unstable systems close to the bifurcation point \citep{dusek94,sipp07,carini15,meliga12}. The original model was proposed by Landau based on symmetry, scaling and phenomenological arguments \citep{landau52}. In the hydrodynamics literature the model is referred to as the Stuart-Landau equation and was first obtained through energy considerations in \cite{stuart58} and via the separation of time scales in \cite{stuart60,watson60}. In order to transform the center-manifold amplitude equations to a form similar to the Stuart-Landau equation, one may multiply equation~\eqref{numerical_amplitude_eqn_x1_p1} with $x^{*}$, and multiply the the complex-conjugate of equation~\eqref{numerical_amplitude_eqn_x1_p1} with $x$. Adding the two resulting equations together leads to an equation for the squared amplitude $x^{*}x = |x|^{2}$. One may then follow the phenomenological arguments of \cite{landau52} for the evolution of the amplitudes in the case of small linear growth rates. Assuming that the system is close to the bifurcation point and the instability is weak, then the change in the amplitude over one cycle of the fundamental frequency also remains small. In such a scenario the evolution equation for the squared amplitude may be averaged over one cycle to obtain the evolution of the average amplitude. The averaging process causes all the oscillatory terms of the type $(x)^{2},(x^{*})^{2},|x|^{2}(x),|x|^{2}(x^{*})$ \textit{etc.} to (nearly) vanish, and the only remaining terms are those related to the modulus of the amplitude. Representing the averaged modulus as $|A|^{2} = \overline{|x|^{2}}$, one obtains the Landau model for the problem as
\begin{equation}
	\label{landau_eqn_1}
		\dfrac{d|A|^{2}}{dt} = 2\times[0.1976\eta + 0.0726\eta^{2}] |A|^{2} - 0.0051|A|^{4}.
\end{equation}
The arguments presented by Landau correspond to the more formal procedure of multiple time-scale expansion and the averaged amplitude $|A|$ corresponds to the amplitude modulation governed by the slow time-scale equations (usually obtained through the solvability condition).

While this is not undertaken in the original work of \cite{landau52}, a similar procedure may be invoked to obtain a slow drift of the angular frequency. Representing $x = |A|e^{i\theta} = |A|e^{i\phi(t)}e^{\lambda_{c}t}$, where, $\phi(t)$ represents the slowly varying drift in angular frequency and $\lambda_{c}$ the critical eigenvalue of the system. One may substitute this expression into equation~\eqref{numerical_amplitude_eqn_x1_p1} and subtract the complex conjugated equation (corresponding to the equation for $x^{*}$) to obtain an expression for $\dt{\phi}$. This equation again contains oscillatory terms of the type $e^{\lambda_{c}t},e^{-\lambda_{c}t},e^{2\lambda_{c}t},e^{-2\lambda_{c}t}$ \textit{etc}. These may be eliminated by averaging over the fundamental time period of the oscillation, with the assumption that $\phi(t)$ remains nearly constant during one cycle of the oscillation, resulting in the expression,
\begin{equation}
	\label{landau_eqn_2}
	\dfrac{d\phi}{dt} = [0.0698\eta - 0.0478\eta^{2}]  + 0.0080|A|^{2}.
\end{equation}
Combining the two equations in $A = |A|e^{i\phi(t)}$ and using the definition of the derivatives in equations~\eqref{landau_eqn_1} and \eqref{landau_eqn_2} one obtains
\begin{eqnarray}
	\label{stuart_landau_eqn}
	\begin{split}
		\dt{A} 					=& [(0.1976 + i0.0698)\eta + (0.0726 - i0.0478)\eta^{2}] A \\
										& - \frac{1}{2}[0.0051 - i0.0160]|A|^{2}A.
	\end{split}
\end{eqnarray}
This is the amplitude equation or, the Stuart-Landau equation for the case in question. In its normal form the Stuart-Landau equation is represented as $\dt{A} = (\sigma_{r} + i\sigma_{i})A - 1/2(l_{r} + il_{i})|A|^{2}A$, and the correspondence is seen immediately. Equation~\eqref{stuart_landau_eqn} may be derived more formally through the usual procedure of multiple time-scale expansion. This however is not undertaken and the phenomenological arguments are deemed sufficient, especially given that a reduced system of center-manifold amplitude evolution has already been obtained in equation~\eqref{numerical_amplitude_eqn_x1_p1}. The Landau constant is found to be $l_{r} = 0.0051$, however, this constant depends on the normalization of the global mode. The ratio $l_{i}/l_{r} = -3.14$ is independent of the normalization. There appears to be some amount of scatter in the reported values of this ratio in the literature, with reported value being $-2.7$  in \cite{dusek94}, $-3.42$ in \cite{sipp07}, $-3.3$ in \cite{meliga12}, and $-3.63$ in \cite{carini15}. The value found in the current work seems to be within the scatter reported in the literature and the reasons behind the scatter are not entirely clear. The ``growth-rate'' of the amplitude that varies linearly with $\eta$ is found here to be $\sigma_{r}=0.1976$ which at first glance appears to be very different from the values reported in \cite{sipp07,meliga12,carini15}. However, in the current work $\eta$ is scaled with $\invRec$  and upon rescaling its value is $9.1488$, which matches well with the values reported in the literature.

%% file: cavity.tex
\section{Flow in an Open Cavity}
\label{application_cavity}

\subsection{The Extended Problem}

\begin{figure}
	\centering
	\includegraphics[width=0.49\textwidth]{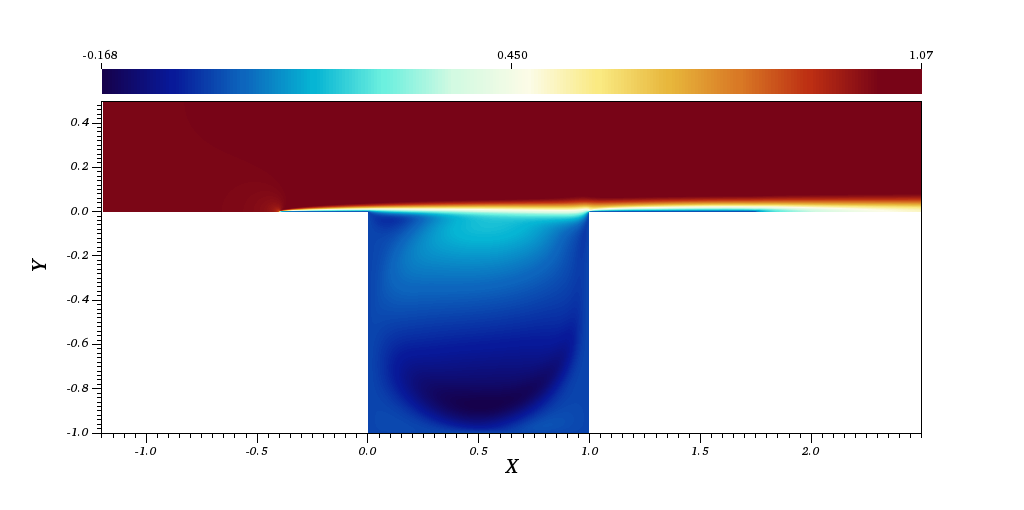}
	\includegraphics[width=0.49\textwidth]{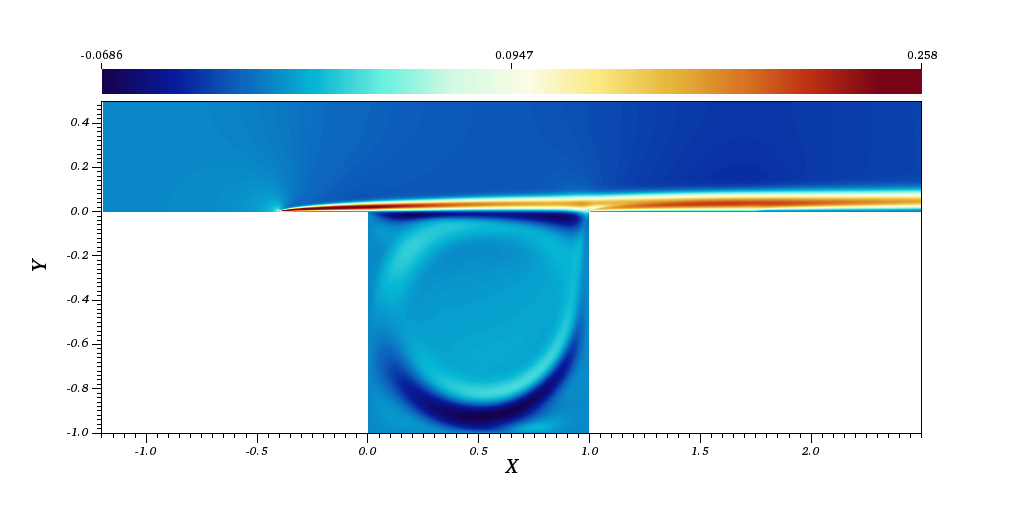}
	\caption{Streamwise velocity of (top) the stationary base flow for the open cavity flow at $\Rey_{c}=4131.33$ and (bottom) the parameter mode.}
	\label{fig:cavity_base_param}
\end{figure}

\begin{figure}
	\centering
	\includegraphics[width=0.49\textwidth]{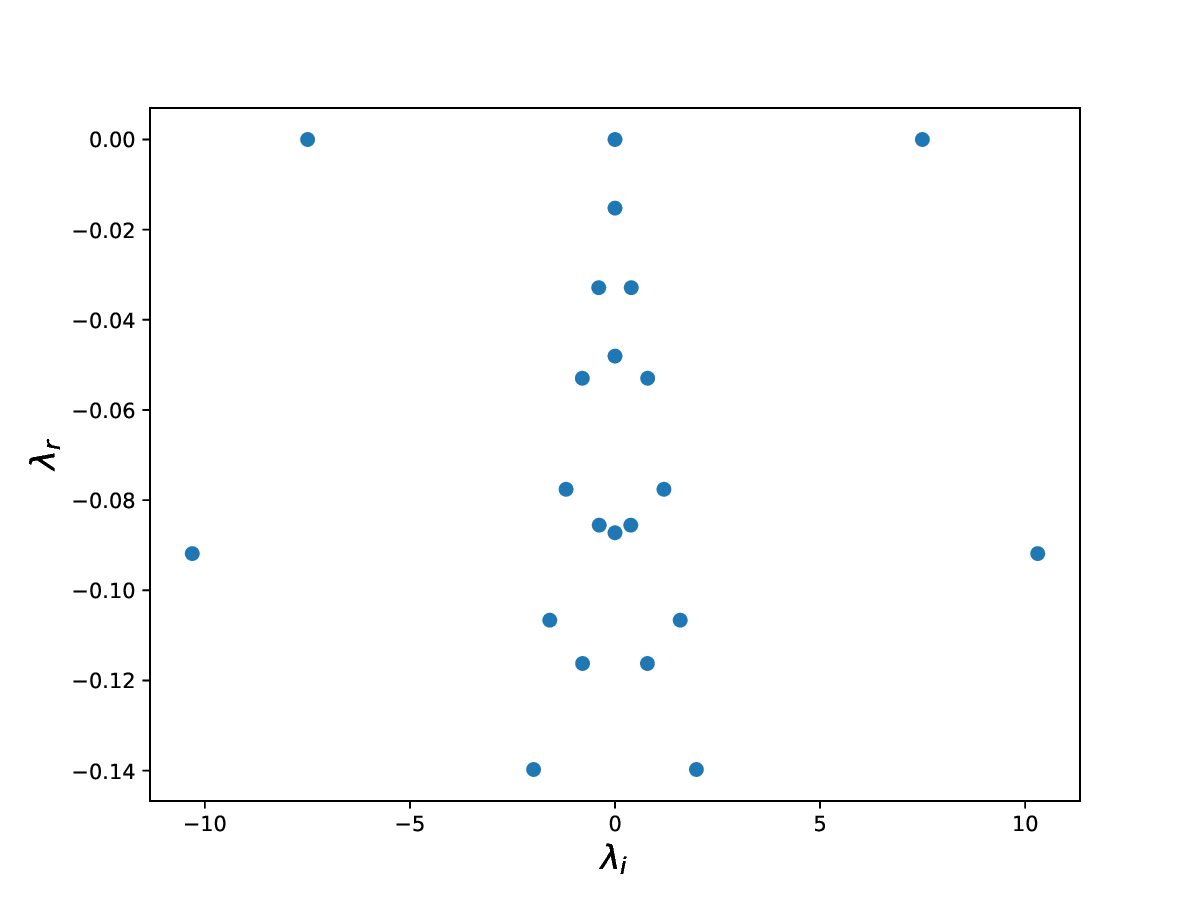}
	\caption{Spectrum of the open cavity flow at the critical point. Note, this is the spectrum of the extended problem which includes the parameter mode located at the origin.}
	\label{fig:cavity_spectrum}
\end{figure}

The case of flow in an open cavity is investigated through the lens of the theoretical results developed in section~\ref{center_manifold_derivation}. This was previously studied by \cite{sipp07} in the context of examining the relevance of mean flow stability analysis to frequency prediction and also by \cite{meliga17} for second-order self-consistent approximations. The problem geometry is set up following \cite{sipp07} and is also shown in figure~\ref{fig:cavity_base_param}. The two dimensional computational domain consists of a square cavity with sides of length one and an open channel flow is constructed above the cavity. The top left corner of the cavity is located at $(x,y)=(0,0)$. The open channel has a width of $0.5$ above the cavity and a symmetry boundary condition is applied to the upper boundary of the channel. The inlet of the channel is located at $x=-1.2$ and a uniform inlet veolcity of $u=1.0$ is applied to the inlet. 
A symmetry boundary condition is applied to the lower wall of the channel from $x=-1.2$ to $x=-0.4$ to allow the flow to develop freely from the inlet. A no-slip condition is applied to the lower walls from $x=-0.4$ to $x=1.75$ (including the cavity walls). Thereafter a symmetry condition is applied again at the lower wall from $x=1.75$ to the outlet, located at $x=2.50$. The critical point for the particular geometry is found to be $Re_{c}=4131.33$, which is close to the values of $4140$ and $4114$ reported in \cite{sipp07} and \cite{meliga17} respectively. The spectrum of the flow at the critical point is shown in figure~\ref{fig:cavity_spectrum}. 

\begin{figure}
	\includegraphics[width=0.49\textwidth]{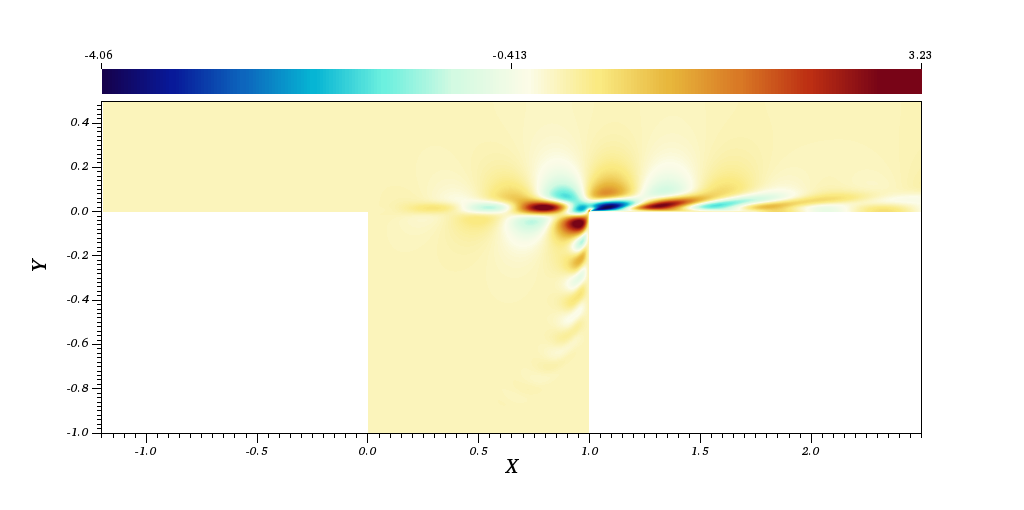}
	\includegraphics[width=0.49\textwidth]{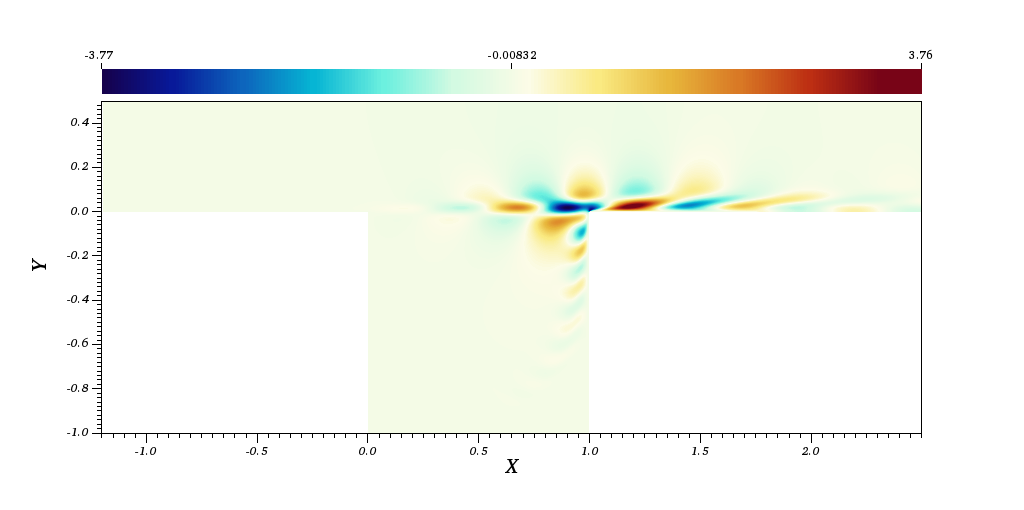}
	\caption{Streamwise velocity of the direct eigenmode. The panels represent the (top) real part and (bottom) imaginary part of the direct critical mode.}
	\label{fig:cavity_flowconfig}
\end{figure}

\begin{figure}
	\includegraphics[width=0.49\textwidth]{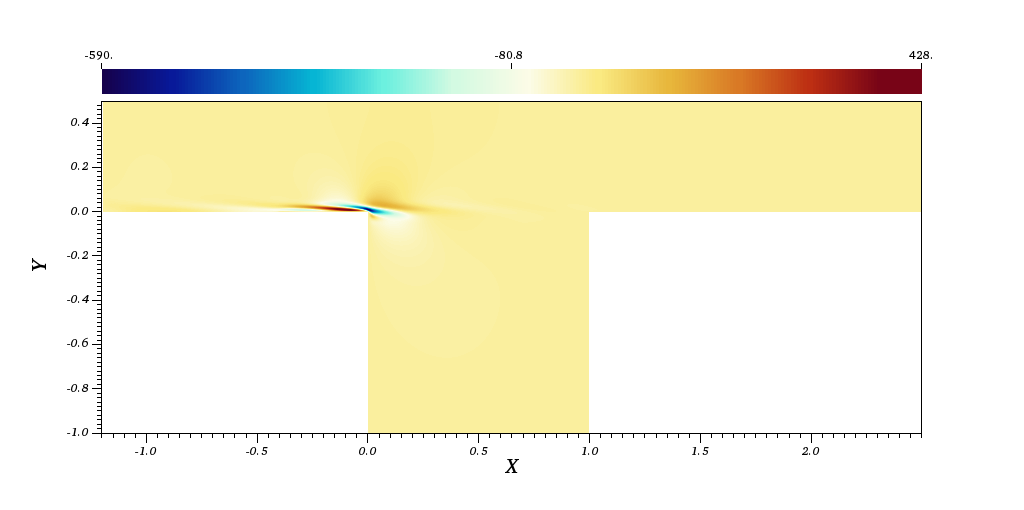}
	\includegraphics[width=0.49\textwidth]{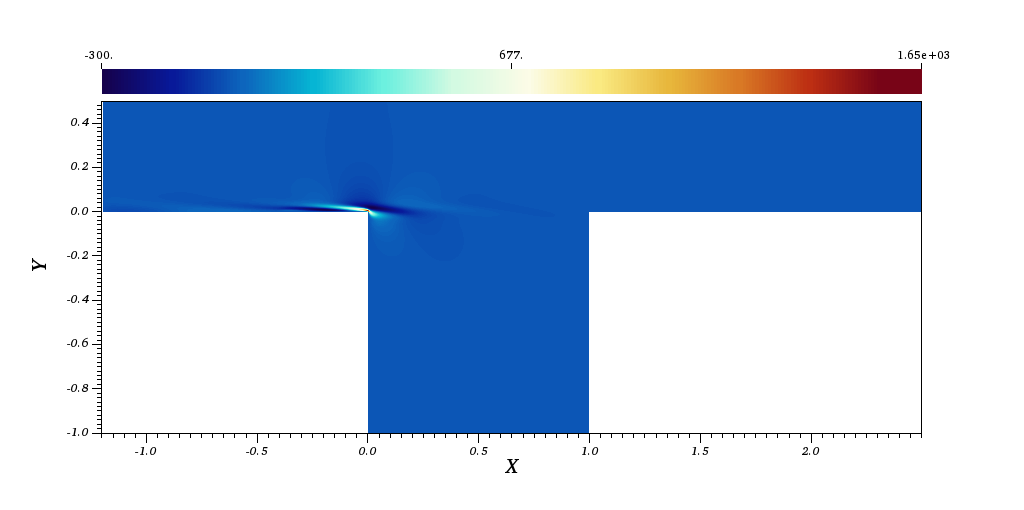}
	\caption{Streamwise velocity of the adjoint eigenmode. The panels represent the (top) real part and (bottom) imaginary part of the adjoint critical mode.}
	\label{fig:cavity_flowconfig_2}
\end{figure}

The streamwise velocity for the calculated baseflow along with the parameter mode is depicted in figure~\ref{fig:cavity_base_param} . The complex conjugate pair of critical eigenvalues for the baseflow is found to be $\lambda_{c} = \pm7.495i$, which compares well with $\lambda_{c}=\pm7.5i$ reported in \cite{sipp07}. The direct and adjoint critical eigenmodes of the flow are shown in figures~\ref{fig:cavity_flowconfig} and ~\ref{fig:cavity_flowconfig_2} respectively. The complex-conjugated mode is not shown. The modes are normalized as earlier with the direct modes being of unit norm and the adjoint modes are scaled appropriately to maintain biorthogonality. Again, the parameter mode is normalized such that $\zeta^{\dagger}_{0} = \zeta_{0} = 1$.

\subsection{Asymptotic Center Manifold}

\begin{figure}
	\centering
	\includegraphics[width=0.49\textwidth]{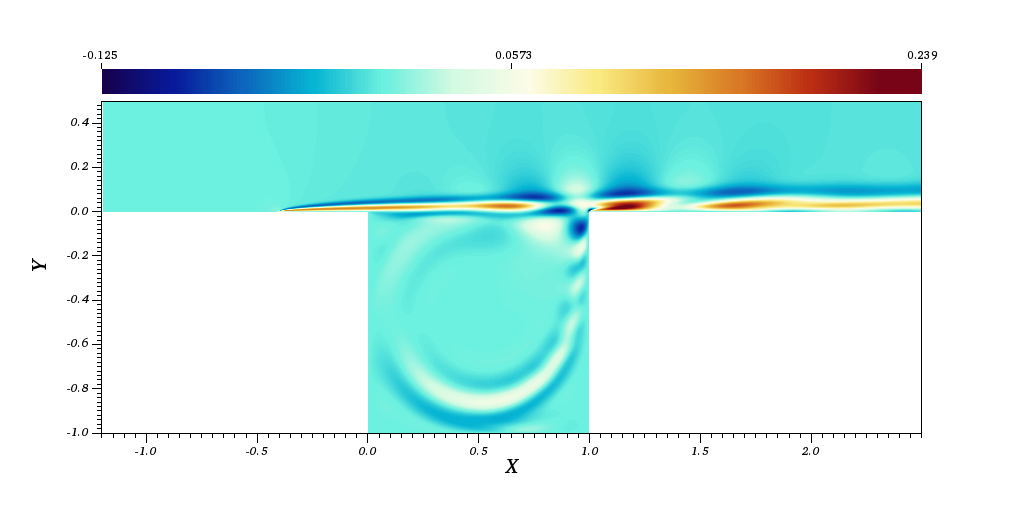}
	\includegraphics[width=0.49\textwidth]{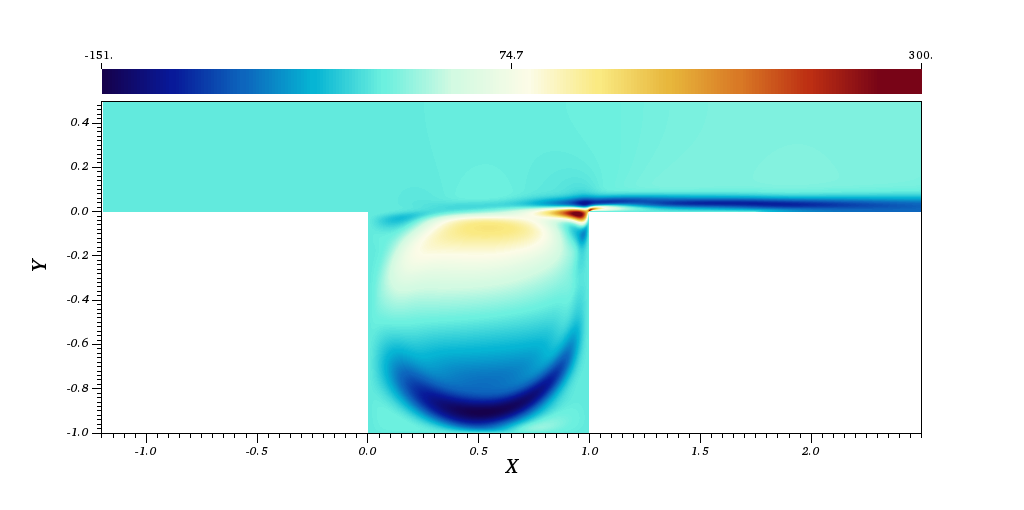}
	\caption{Streamwise velocity components of the restricted resolvent solutions for the open cavity flow corresponding to the zero frequency fields (top) $\vpfield{y}_{0,0}$ and (bottom) $\vpfield{y}_{1,2}$ for the open cavity flow case.}
	\label{fig:cavity_resolvents_zero}
\end{figure}

\begin{figure}
	\centering
	\includegraphics[width=0.49\textwidth]{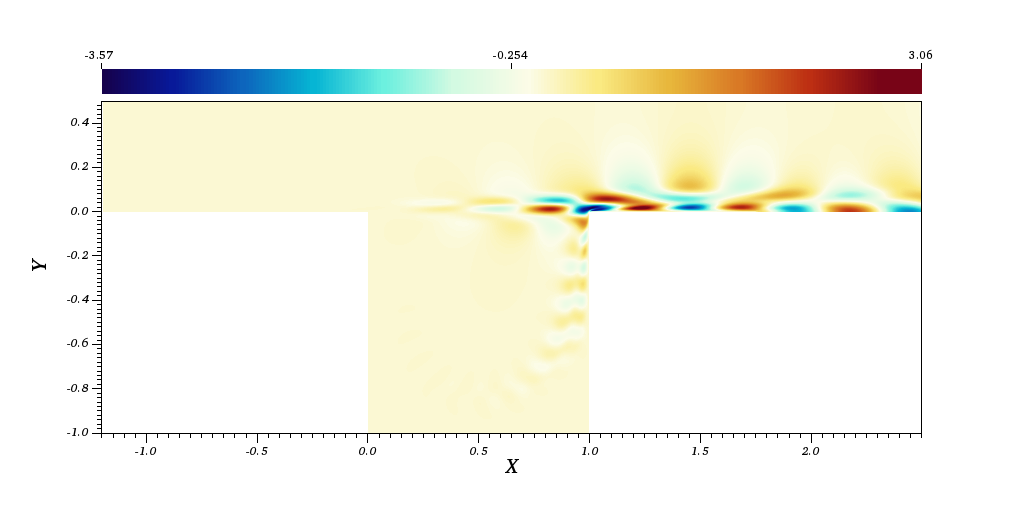}
	\includegraphics[width=0.49\textwidth]{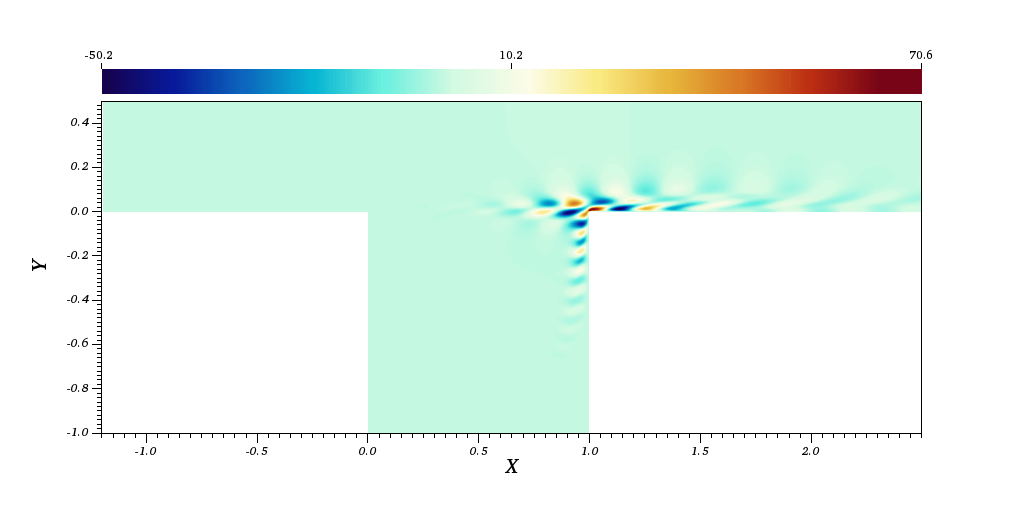}
	\caption{The panels depict the real part of the streamwise velocity components of (top) $\vpfield{y}_{0,1}$ and (bottom) $\vpfield{y}_{1,1}$ with angular frequencies equal to the imaginary parts of $\lambda_{c}$ and $2\lambda_{c}$.}
	\label{fig:cavity_resolvents_oscillatory_real}
\end{figure}

\begin{figure}
	\centering
	\includegraphics[width=0.49\textwidth]{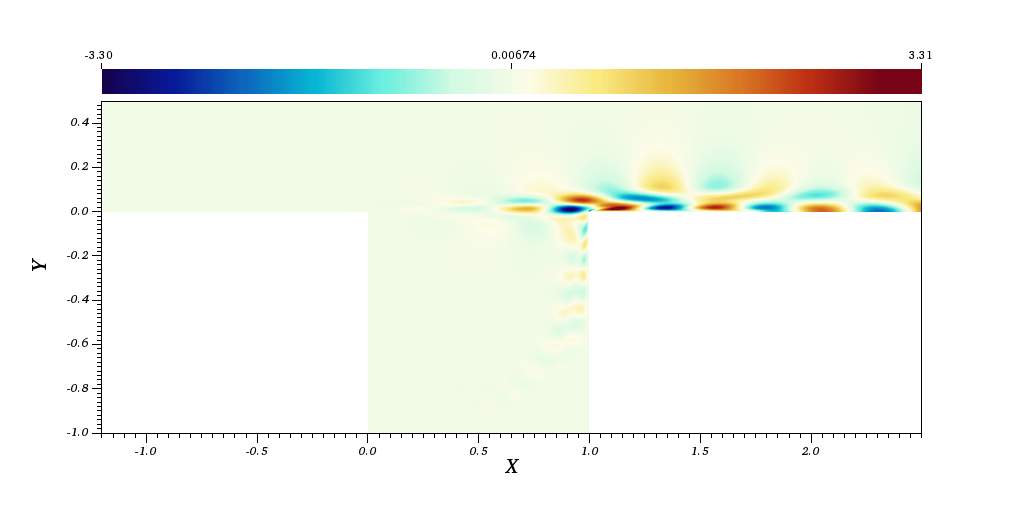}
	\includegraphics[width=0.49\textwidth]{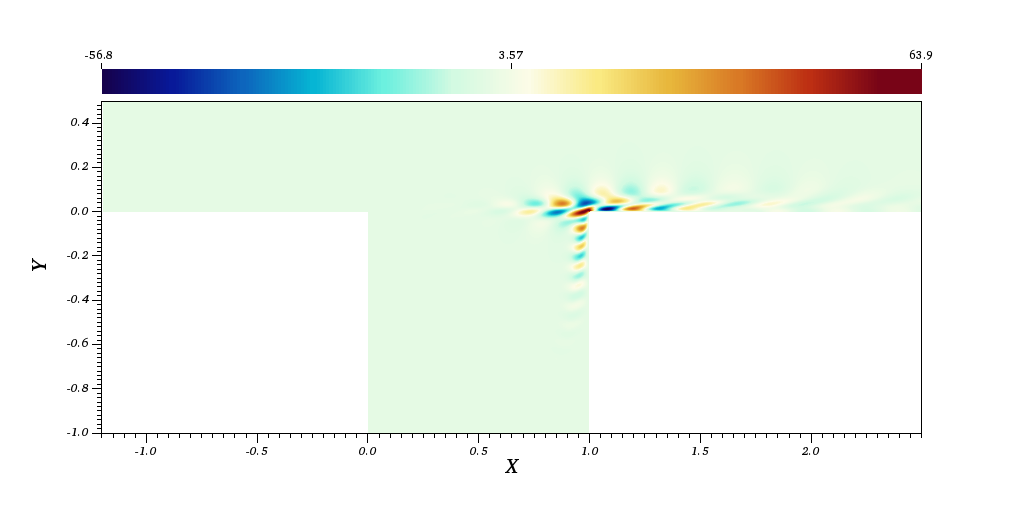}
	\caption{The panels depict the imaginary part of the streamwise velocity components of (top) $\vpfield{y}_{0,1}$ and (bottom) $\vpfield{y}_{1,1}$ with angular frequencies equal to the imaginary parts of $\lambda_{c}$ and $2\lambda_{c}$.}
	\label{fig:cavity_resolvents_oscillatory_imag}
\end{figure}

As done earlier, second order asymptotic approximation of the graph $\mathfrak{h}(\mathsfbi{x})$ are evalued which requires the evaluation of six restricted resolvent solution fields $\vpfield{y}_{i,j}$. The zero frequency fields that contribute to the mean flow correction ($\vpfield{y}_{0,0}$ and $\vpfield{y}_{1,2}$) are depicted in figure~\ref{fig:cavity_resolvents_zero} while the oscillatory resolvent fields $\vpfield{y}_{0,1}$ and $\vpfield{y}_{1,1}$ are depicted in figures~\ref{fig:cavity_resolvents_oscillatory_real} and \ref{fig:cavity_resolvents_oscillatory_imag}. The corresponding complex conjugated fields ($\vpfield{y}_{0,2}$ and $\vpfield{y}_{2,2}$) are not displayed.


 The center-manifold amplitude equation can be easily obtained once the resolvent fields have been calculated.  Replacing $x_{0},x_{1},x_{2}$ with $\eta,x,x^{*}$, one obtains the following equation for $\dt{x}$,
\begin{equation}
	\label{cavity_numerical_amplitude_eqn_x1_p1}
	\begin{aligned}
			\dt{x} =& +[7.495i]x \\
			& + [(11.72 - 6.618i)\times10^{-2}]\eta^{2} \\
			& + [(8.345 + 7.238i)\times10^{-1}]\eta x \\
			 & - [(1.544 - 5.471i)\times10^{-2}]\eta x^{*}  \\
			& + [(18.42 - 4.701i)\times10^{-1}] x^{2} \\
			 & + [(2.268 - 3.955i)\times10^{+0}] x^{*}x \\
			& + [(1.007 + 4.994i)\times10^{-1}](x^{*})^{2} \\
			& + [(9.656 - 5.478i)\times10^{-2}]\eta^{3} \\
			& + [(3.188 + 2.514i)\times10^{-1}]\eta^{2}x \\
			 & - [(7.922 - 2.635i)\times10^{-2}]\eta^{2}x^{*}  \\
			& + [(12.94 - 6.695i)\times10^{-1}]\eta x^{2}  \\
			 & + [(0.898 - 1.946i)\times10^{+1}]\eta x^{*}x \\
			& + [(2.234 - 2.946i)\times10^{-1}]\eta(x^{*})^{2} \\
			 & - [(3.701 - 3.134i)\times10^{+0}]x^{3} \\
			& - [(5.744 - 3.425i)\times10^{+2}](x^{*}x)x \\
			& - [(9.844 - 3.445i)\times10^{+1}](x^{*}x)x^{*} \\
			& - [(4.613 - 6.681i)\times10^{+0}](x^{*})^{3}.
		\end{aligned}
\end{equation}

\begin{figure}
	\centering
	\includegraphics[width=.49\textwidth]{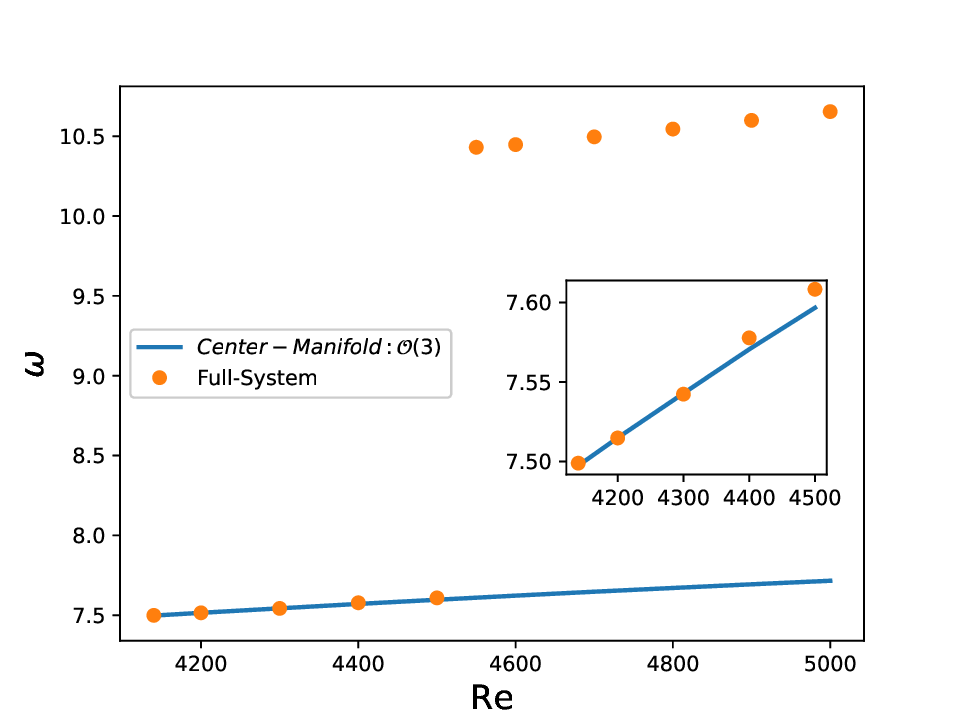}
	\caption{Comparison of the non-linear angular frequencies of the full system and the center-manifold amplitude equations for the open cavity flow. The frequencies before the mode switching are in good agreement with each other.}
	\label{fig:cavity_limit_cycle_frequency}
\end{figure}

In contrast to the reduced equation obtained for the flow across a cylinder, there are no obvious symmetries in the problem and therefore extremely small terms are not obtained for the reduced equations of the open cavity flow. Additionally, the center-manifold amplitude equations are now inhomogeneous, since the terms of the type $\eta^{2}, \eta^{3}$ now have a non-zero coefficient.  The comparison of the angular frequencies obtained through the center manifold equations with those of the full system is shown in figure~\ref{fig:cavity_limit_cycle_frequency}. The system has an interesting behavior past the bifurcation point wherein, initially the system evolves with an angular frequency close to that of the critical mode at bifurcation, clearly indicating the continuation of the system dynamics along the center manifold. However, around $Re=4550$ the non-linear frequencies change abruptly. The spectrum at bifurcation shown in figure~\ref{fig:cavity_spectrum} indicates that the new angular frequency is likely related to the isolated subcritical mode of the system, implying the occurence of mode switching. Obviously the mode switching behaviour can not be captured by the center-manifold amplitude equations since the information regarding the isolated subcritical mode is not captured by the reduced system of equations. Nonetheless, before the mode switching occurs, the angular frequency predicted by the reduced equations agrees very well with those obtained from the full system. 

The equation for the Landau model and the slow phase variation are obtained as 
\begin{align}
	\label{cavity_landau_eqn_1}
		\dfrac{d|A|^{2}}{dt} =& 2\times[0.834\eta + 0.319\eta^{2}] |A|^{2} - 1148.78|A|^{4}, \\
	\label{cavity_landau_eqn_2}
	\dfrac{d\phi}{dt} =& [0.7238\eta + 0.2514\eta^{2}]  + 342.47|A|^{2},
\end{align}
with the combined Stuart-Landau equation being
\begin{eqnarray}
	\label{cavity_stuart_landau_eqn}
	\begin{split}
		\dt{A} 					=& [(0.834 + i0.7238)\eta + (0.319 + i0.2514)\eta^{2}] A \\
										& - \frac{1}{2}[1148.78 - i684.94]|A|^{2}A.
	\end{split}
\end{eqnarray}

This presents a rather interesting case for the mathematical modeling of fluid flows since the mode switching behavior is not captured by the center-manifold nevertheless the flow has a distinct angular frequency and no sign of chaotic behavior clearly suggesting the evolution of the dynamics in some low dimensional subspace. The relaxation of the assumption of critical eigenvalues of the dynamic subspace leads naturally to the invariant manifold theory however, the application of this theory requires the existence of a spectral gap between the eigenvalues of the dynamic and driven subspaces \citep{chicone06}. The spectrum plotted in figure~\ref{fig:cavity_spectrum} indicates that if one considers the eigenmodes corresponding to the two isolated eigenvalues as part of the dynamic subspace the spectral gap assumption is not fullfilled. The mathematical modeling in such a scenario clearly requires a further refinement of the underlying assumptions. 

Finally, a few comments may be made with regards to the comparison of the current work with the weakly non-linear analysis that is common in hydrodynamics. Despite claims of equivalence \citep{fujimara91} of the final Landau equations, the formal procedure for evaluation of solution is quite different for the two methods. Nonetheless, given that both methods are applied close to the bifurcation point and are trying to capture essentially the same physics, one would expect that the results of the two methods would indeed be similar. Indeed some of the coefficients of the Stuart-Landau equation obtained for the case of the cylinder are similar to those reported in the literature using weakly nonlinear analysis \citep{sipp07}. However, some differences do seem to arise. In the current work, the coefficient of the linear term $A$ is obtained up to quadratic order in $\eta$. The author is unaware of any work utilizing weakly non-linear analysis for the current cases that reports such a quadratic variation. Differences also arise in the coefficient of the cubic term $|A|^{2}A$. The work of \cite{sipp07} contains a linear dependence on the parameter for this term while in the current investigation the cubic saturation term is devoid of any such dependence. Formally, a term of the type $\eta|A|^{2}A$, when viewed in the current framework would be a fourth order term since the parameter perturbation is treated through the term $x_{0}$ which is just another mode amplitude. These are structural differences in the solutions which no doubt partly arise due to the distinct treatment of the parameter variation. Such structural differences in the two methods could be an example of the claims made in \cite{fujimara91} that high order evaluations may be required to achieve exact equivalences. A systematic comparison, including evaluation of higher orders, would be required to tease out such differences.

%% file: conclusion.tex
\section{Conclusion}
\label{conclusion}
A center-manifold reduction for the Navier--Stokes equation is derived with the extension of the system to include the trivial parameter evolution equation. This inclusion however enlarges the critical subspace in a non-trivial manner. The system is then reformulated to bring it to an appropriate form and the application of center-manifold theorem leads to a graph equation for the reformulated system. The graph equation is a representation of the infinite dimensional stable subspace that is driven by the dynamics in the critical subspace. The equation is solved asymptotically and expressions for the asymptotic solutions are obtained, which finally result in a set of equations for center-manifold amplitude equations for the critical eigenvectors. The derivation is essentially agnostic to the dimension of the critical subspace however, the number terms required for the asymptotic evaluation rises rapidly with increasing critical subspace dimension. While the derivation is set in the context of the Navier--Stokes, the methodology is more general and equations~\eqref{general_second_order_expression} and \eqref{general_higher_order_expression} point to the essential structure of the asymptotic solutions of the graph equation that may be obtained for other infinite-dimensional problems. 

The proposed method with system extension is of particular relevance when considering center-manifold reduction for problems that have been perturbed away from the bifurcation point and is a formal way to incorporate parameter perturbations within the asymptotic approximations of the graph equation thereby negating the need for a double asymptotic expansion or an apriori assumption of the normal form of the reduced dynamics.

The derivation is then applied to two cases - the case of the Hopf bifurcation of a cylinder wake and flow in an open cavity resulting in the reduced center-manifold amplitude equations from which, the Stuart-Landau equations are derived. The results from the reduced equations provide a good prediction for the frequencies of the full system close to the bifurcation point. In particular, the linearized eigenvalue variation is predicted up to fourth order in $\eta$ for the case of the cylinder flow and it is found to match well with the spectra calculation of the full system at different Reynolds numbers. The case of flow in an open cavity has interesting dynamics with clear evidence of mode switching at a certain parameter value. This mode switching behavior can not be captured within the bounds of center-manifold theory. Nevertheless, the frequency prior to the mode switching is indeed predicted well by the center-manifold reduction of the extended system.

\section*{Acknowledgements}
The author is grateful for the computational resources and support provided by the Scientific Computing and Data Analysis section of Research Support Division at OIST.

The research was supported by the Okinawa Institute of Science and Technology Graduate	University (OIST) with subsidy funding from the Cabinet Office, Government of Japan. Part of the research was conducted at the Nordic Institute for Theoretical Physics, Nordita, and the author acknowledges the support of Nordita and the Swedish Research Council Grant No. 2018-04290.

\FloatBarrier

%% file: appendix.tex
\appendix

\section{Center-Manifold Coefficients} 
\label{AppendixA}

Table~\ref{tab:coeffs_x1} reports the evaluated coefficients for the different polynomial terms in the center-manifold amplitude equations for the Hopf bifurcations in the cylinder wake and for the flow in an open cavity. The reported coefficient are for $\dt{x}$ and one obtains the complex conjugated terms for $\dt{x}^{*}$ up to an accurary of $10^{-8}$, where numerical accuracy and solver tolerances cause a violation of the complex conjugation property.

\begin{table*}
	\centering
	\caption{Coefficients of polynomial terms for $\dot{x}_{1}$ obtained numerically for the two cases. }
	\begin{tabular}{c | c | c}
		Polynomial Terms  & Cylinder Wake		                  & Open Cavity\\
		\hline\hline
		$x_{0}$           & $0.0 + 0.0i$ 				      & $0.0 + 0.0i$ 			     \\
		$x_{1}$           & $0.0 + 0.7455i$ 				& $0.0 + 7.495i$   		     \\
		$x_{2}$           & $0.0 + 0.0i$ 					& $0.0 + 0.0i$ 			     \\
            $x_{0}x_{0}$      & $(+3.640 + 2.291i)\times10^{-8}$ 	      & $(+11.72 - 6.618i)\times10^{-2}$ \\
		$x_{0}x_{1}$      & $(+1.976 + 0.698i)\times10^{-1}$ 		& $(+8.345 + 7.238i)\times10^{-1}$ \\
		$x_{0}x_{2}$      & $(+6.388 + 4.616i)\times10^{-3}$ 		& $(-1.544 + 5.471i)\times10^{-2}$ \\
		$x_{1}x_{1}$      & $(-1.479 - 0.211i)\times10^{-11}$ 	& $(+18.42 - 4.701i)\times10^{-1}$ \\
		$x_{1}x_{2}$      & $(-1.306 + 1.290i)\times10^{-10}$ 	& $(+2.268 - 3.955i)\times10^{+0}$ \\
            $x_{2}x_{2}$      & $(+0.380 - 1.652i)\times10^{-11}$ 	& $(+1.007 + 4.994i)\times10^{-1}$ \\
            $x_{0}x_{0}x_{0}$	& $(+2.954 + 1.373i)\times10^{-8}$        & $(+9.656 - 5.478i)\times10^{-2}$ \\ 
            $x_{0}x_{0}x_{1}$	& $(+7.264 - 4.787i)\times10^{-2}$        & $(+3.188 + 2.514i)\times10^{-1}$ \\
            $x_{0}x_{0}x_{2}$ & $(+4.477 - 1.482i)\times10^{-3}$        & $(-7.922 + 2.635i)\times10^{-2}$ \\
            $x_{0}x_{1}x_{1}$ & $(-3.854 - 3.090i)\times10^{-9}$        & $(+12.94 - 6.695i)\times10^{-1}$ \\
            $x_{0}x_{1}x_{2}$ & $(+0.711 + 2.542i)\times10^{-9}$        & $(+0.898 - 1.946i)\times10^{+1}$ \\
            $x_{0}x_{2}x_{2}$ & $(+1.077 + 0.492i)\times10^{-10}$       & $(+2.234 - 2.946i)\times10^{-1}$ \\
            $x_{1}x_{1}x_{1}$ & $(+1.615 - 2.904i)\times10^{-5}$        & $(-3.701 + 3.134i)\times10^{+0}$ \\
            $x_{1}x_{1}x_{2}$ & $(-2.526 + 8.010i)\times10^{-3}$        & $(-5.744 + 3.425i)\times10^{+2}$ \\
            $x_{1}x_{2}x_{2}$ & $(+0.973 - 3.380i)\times10^{-5}$        & $(-9.844 + 3.445i)\times10^{+1}$ \\
            $x_{2}x_{2}x_{2}$ & $(-2.308 -5.938i)\times10^{-7}$        & $(-4.613 + 6.681i)\times10^{+0}$ \\
		\hline\hline
	\end{tabular} 
	\label{tab:coeffs_x1}
\end{table*}


\section{Bifurcation Algorithm}
\label{AppendixB}

The following describes the numerical procedure for obtaining the bifurcation point. 
First three Reynolds numbers are selected, representing the parameter range and its mid-point for the search of the critical Reynolds number. Maximum growth rates $\sigma$ at the three points are determined using the Krylov-Schur method \citep{stewart02} for the spectral problem. Using the three points a quadratic relation is built for the Reynolds number as a function of the growth rate \ie, $\Rey(\sigma) = a + b\sigma + c\sigma^{2}$. The coefficients $a,b,c$, obtained through the spectral results obtained at the three points. An approximate prediction of the critical point is found using $\Rey(0) = a$. The fixed point solution and the spectral poblem is solved again to obtain $\sigma$ at the predicted $\Rey$. If the newly obtained growth rate $\sigma$ is greater than a specified tolerance, a new relation $Re(\sigma)$ is built using the smallest three growth rate values and the next prediction of the critical point is obtained.  The procedure is repeated till a required tolerance for the critical point is obtained. In the current work the tolerance of $\sigma<10^{-9}$ is used for the determination of the critical point, and the critical Reynolds number is found to be $\Rey_{c} = 46.30$ for the case of flow across a stationary cylinder and $\Rey_{c}=4131.33$ for the open cavity.

%% file: main.bbl
\begin{thebibliography}{}

\bibitem[{\AA}kervik et~al., 2006]{akervik06}
{\AA}kervik, E., Brandt, L., Henningson, D.~S., Hœpffner, J., Marxen, O., and
  Schlatter, P. (2006).
\newblock Steady solutions of the {N}avier-{S}tokes equations by selective
  frequency damping.
\newblock {\em Physics of Fluids}, 18(6):068102.

\bibitem[Barkley, 2006]{barkley06}
Barkley, D. (2006).
\newblock Linear analysis of the cylinder wake mean flow.
\newblock {\em Europhysics Letters ({EPL})}, 75(5):750--756.

\bibitem[Bender and Orzag, 1999]{bender99}
Bender, C.~M. and Orzag, S.~A. (1999).
\newblock {\em Advanced Mathematical Methods for Scientists and Engineers}.
\newblock Springer Science \& Business Media New York.

\bibitem[Carini et~al., 2015]{carini15}
Carini, M., Auteri, F., and Giannetti, F. (2015).
\newblock Centre-manifold reduction of bifurcating flows.
\newblock {\em Journal of Fluid Mechanics}, 767:109–145.

\bibitem[Carr, 1982]{carr82}
Carr, J. (1982).
\newblock {\em Applications of centre manifold theory}, volume~35.
\newblock Springer Science \& Business Media.

\bibitem[Carr and Muncaster, 1983a]{carr83b}
Carr, J. and Muncaster, R.~G. (1983a).
\newblock The application of centre manifolds to amplitude expansions. i.
  ordinary differential equations.
\newblock {\em Journal of Differential Equations}, 50(2):260--279.

\bibitem[Carr and Muncaster, 1983b]{carr83}
Carr, J. and Muncaster, R.~G. (1983b).
\newblock The application of centre manifolds to amplitude expansions. ii.
  infinite dimensional problems.
\newblock {\em Journal of differential equations}, 50(2):280--288.

\bibitem[Chicone, 2006]{chicone06}
Chicone, C.~C. (2006).
\newblock {\em Ordinary differential equations with applications}, volume~34.
\newblock Springer.

\bibitem[Coullet and Spiegel, 1983]{coullet83}
Coullet, P.~H. and Spiegel, E.~A. (1983).
\newblock Amplitude equations for systems with competing instabilities.
\newblock {\em SIAM Journal on Applied Mathematics}, 43(4):776--821.

\bibitem[Cross and Greenside, 2009]{cross09}
Cross, M. and Greenside, H. (2009).
\newblock {\em Pattern Formation and Dynamics in Nonequilibrium Systems}.
\newblock Cambridge University Press.

\bibitem[Deville et~al., 2002]{deville02}
Deville, M.~O., Fischer, P.~F., and Mund, E.~H. (2002).
\newblock {\em High-Order Methods for Incompressible Fluid Flow}.
\newblock Number~9 in Cambridge Monographs on Applied and Computational
  Mathematics. Cambridge University Press.

\bibitem[Dušek et~al., 1994]{dusek94}
Dušek, J., Gal, P.~L., and Fraunié, P. (1994).
\newblock A numerical and theoretical study of the first {H}opf bifurcation in
  a cylinder wake.
\newblock {\em Journal of Fluid Mechanics}, 264:59–80.

\bibitem[Fischer et~al., 2008]{nek5000}
Fischer, P.~F., Lottes, J.~W., and Kerkemeier, S.~G. (2008).
\newblock Nek5000 web page.
\newblock \url{http://nek5000.mcs.anl.gov}.

\bibitem[Fujimura, 1991]{fujimara91}
Fujimura, K. (1991).
\newblock Methods of centre manifold and multiple scales in the theory of
  weakly nonlinear stability for fluid motions.
\newblock {\em Proceedings: Mathematical and Physical Sciences},
  434(1892):719--733.

\bibitem[Fujimura, 1997]{fujimura97}
Fujimura, K. (1997).
\newblock Centre manifold reduction and the {S}tuart-{L}andau equation for
  fluid motions.
\newblock {\em Proceedings of the Royal Society of London. Series A:
  Mathematical, Physical and Engineering Sciences}, 453(1956):181--203.

\bibitem[Gianetti and Luchini, 2007]{giannetti07}
Gianetti, F. and Luchini, P. (2007).
\newblock Structural sensitivity of the first instability of the cylinder wake.
\newblock {\em Journal of Fluid Mechanics}, 581:167–197.

\bibitem[Guckenheimer and Holmes, 1983]{guckenheimer83}
Guckenheimer, J. and Holmes, P. (1983).
\newblock {\em Nonlinear Oscillations, Dynamical Systems, and Bifurcations of
  Vector Fields}.
\newblock Springer Science \& Business Media New York.

\bibitem[Guckenheimer and Knobloch, 1983]{guckenheimer83b}
Guckenheimer, J. and Knobloch, E. (1983).
\newblock Nonlinear convection in a rotating layer: amplitude expansions and
  normal forms.
\newblock {\em Geophysical \& Astrophysical Fluid Dynamics}, 23(4):247--272.

\bibitem[Haragus and G\'{e}rard, 2011]{haragus11}
Haragus, M. and G\'{e}rard, I. (2011).
\newblock {\em Local Bifurcations, Center Manifolds, and Normal Forms in
  Infinite-Dimensional Dynamical Systems}.
\newblock Springer-Verlag New York.

\bibitem[Knobloch and Guckenheimer, 1983]{knobloch83}
Knobloch, E. and Guckenheimer, J. (1983).
\newblock Convective transitions induced by a varying aspect ratio.
\newblock {\em Physical Review A}, 27(1):408.

\bibitem[Landau, 1944]{landau52}
Landau, L.~D. (1944).
\newblock On the problem of turbulence.
\newblock In HAAR, D.~T., editor, {\em Collected Papers of L.D. Landau}, pages
  387--391. Pergamon.

\bibitem[Maday and Patera, 1989]{maday89}
Maday, Y. and Patera, A.~T. (1989).
\newblock Spectral element methods for the incompressible {N}avier-{S}tokes
  equations.
\newblock In {\em State-of-the-art surveys on computational mechanics}, pages
  71--143. American Society of Mechanical Engineers, New York.

\bibitem[Manti{\v{c}}-Lugo et~al., 2015]{mantivc15}
Manti{\v{c}}-Lugo, V., Arratia, C., and Gallaire, F. (2015).
\newblock A self-consistent model for the saturation dynamics of the vortex
  shedding around the mean flow in the unstable cylinder wake.
\newblock {\em Physics of Fluids}, 27(7).

\bibitem[Meliga, 2017]{meliga17}
Meliga, P. (2017).
\newblock Harmonics generation and the mechanics of saturation in flow over an
  open cavity: a second-order self-consistent description.
\newblock {\em Journal of Fluid Mechanics}, 826:503--521.

\bibitem[Meliga and Chomaz, 2011]{meliga11}
Meliga, P. and Chomaz, J.-M. (2011).
\newblock An asymptotic expansion for the vortex-induced vibrations of a
  circular cylinder.
\newblock {\em Journal of Fluid Mechanics}, 671:137–167.

\bibitem[Meliga et~al., 2012]{meliga12}
Meliga, P., Gallaire, F., and Chomaz, J.-M. (2012).
\newblock A weakly nonlinear mechanism for mode selection in swirling jets.
\newblock {\em Journal of Fluid Mechanics}, 699:216–262.

\bibitem[Negi, 2019]{negiphd}
Negi, P. (2019).
\newblock {\em Stability and transition in pitching wings}.
\newblock PhD thesis, KTH Royal Institute of Technology.

\bibitem[Negi et~al., 2020]{negi20}
Negi, P.~S., Hanifi, A., and Henningson, D.~S. (2020).
\newblock On the linear global stability analysis of rigid-body motion
  fluid–structure-interaction problems.
\newblock {\em Journal of Fluid Mechanics}, 903:A35.

\bibitem[Newell and Whitehead, 1969]{newell69}
Newell, A.~C. and Whitehead, J.~A. (1969).
\newblock Finite bandwidth, finite amplitude convection.
\newblock {\em Journal of Fluid Mechanics}, 38(2):279–303.

\bibitem[Patera, 1984]{patera84}
Patera, A.~T. (1984).
\newblock A spectral element method for fluid dynamics: Laminar flow in a
  channel expansion.
\newblock {\em Journal of Computational Physics}, 54(3):468 -- 488.

\bibitem[Pier, 2002]{pier02}
Pier, B. (2002).
\newblock On the frequency selection of finite-amplitude vortex shedding in the
  cylinder wake.
\newblock {\em Journal of Fluid Mechanics}, 458:407–417.

\bibitem[Provansal et~al., 1987]{provansal87}
Provansal, M., Mathis, C., and Boyer, L. (1987).
\newblock Bénard-von {K}ármán instability: transient and forced regimes.
\newblock {\em Journal of Fluid Mechanics}, 182:1–22.

\bibitem[Roberts, 1997]{roberts97}
Roberts, A. (1997).
\newblock Low-dimensional modelling of dynamics via computer algebra.
\newblock {\em Computer Physics Communications}, 100(3):215--230.

\bibitem[Roberts, 1989]{roberts89}
Roberts, A.~J. (1989).
\newblock The utility of an invariant manifold description of the evolution of
  a dynamical system.
\newblock {\em SIAM Journal on Mathematical Analysis}, 20(6):1447--1458.

\bibitem[Sijbrand, 1985]{sijbrand85}
Sijbrand, J. (1985).
\newblock Properties of center manifolds.
\newblock {\em Transactions of the American Mathematical Society},
  289(2):431--469.

\bibitem[Sipp and Lebedev, 2007]{sipp07}
Sipp, D. and Lebedev, A. (2007).
\newblock Global stability of base and mean flows: a general approach and its
  applications to cylinder and open cavity flows.
\newblock {\em Journal of Fluid Mechanics}, 593:333–358.

\bibitem[Stewart, 2002]{stewart02}
Stewart, G.~W. (2002).
\newblock A {K}rylov--{S}chur algorithm for large eigenproblems.
\newblock {\em SIAM Journal on Matrix Analysis and Applications},
  23(3):601--614.

\bibitem[Stuart, 1958]{stuart58}
Stuart, J.~T. (1958).
\newblock On the non-linear mechanics of hydrodynamic stability.
\newblock {\em Journal of Fluid Mechanics}, 4(1):1–21.

\bibitem[Stuart, 1960]{stuart60}
Stuart, J.~T. (1960).
\newblock On the non-linear mechanics of wave disturbances in stable and
  unstable parallel flows part 1. the basic behaviour in plane poiseuille flow.
\newblock {\em Journal of Fluid Mechanics}, 9(3):353–370.

\bibitem[Watson, 1960]{watson60}
Watson, J. (1960).
\newblock On the non-linear mechanics of wave disturbances in stable and
  unstable parallel flows part 2. the development of a solution for plane
  {P}oiseuille flow and for plane {C}ouette flow.
\newblock {\em Journal of Fluid Mechanics}, 9(3):371–389.

\bibitem[Wiggins, 2003]{wiggins03}
Wiggins, S. (2003).
\newblock {\em Introduction to Applied Nonlinear Dynamical Systems and Chaos}.
\newblock Springer-Verlag New York.

\end{thebibliography}
